\documentclass[fleqn,usenatbib,useAMS]{mnras}

\usepackage{graphicx}	
\usepackage{amsmath}	
\usepackage{amssymb}	
\usepackage{amsfonts}
\usepackage{amsbsy}
\usepackage{color}
\usepackage{rotating}
\usepackage{multicol}        
\usepackage{dcolumn}
\usepackage{bm}		
\usepackage{lscape}	
\usepackage[T1]{fontenc}
\usepackage[english]{babel}
\usepackage{hyperref}
\usepackage{ae,aecompl}
\usepackage{cleveref}
\usepackage[rightcaption]{sidecap}
\usepackage{subfigure}

\usepackage{%
graphicx,%
array,%
bm,%
booktabs,%
multirow%
}
\usepackage[utf8]{inputenc}
\usepackage[english]{babel}
\usepackage{amsmath,amsfonts,amssymb}
\usepackage[usenames,dvipsnames,svgnames]{xcolor}
\usepackage[T1]{fontenc}
\usepackage{capt-of}
\usepackage{tabularx}

\newcommand{\be}{\begin{equation}}
\newcommand{\ee}{\end{equation}}

\renewcommand{\tilde}{\widetilde}

\usepackage{etoolbox}
\makeatletter
\appto{\appendix}{%
  \@ifstar{\def\thesection{\unskip}\def\theequation@prefix{A.}}%
          {\def\thesection{\Alph {section}}}%
}
\makeatother

\Crefrangeformat{equation}{Eqs. (#3#1#4)--(#5#2#6)}
\Crefname{equation}{Eq.}{Eqs.}
\Crefname{figure}{Figure}{Figures}
\Crefname{section}{Section}{Sections}

\title[Cosmographic
 variables and Padé
polynomials]{High-redshift cosmography:
auxiliary variables versus Padé
polynomials}

\author[S.~Capozziello, R.~D'Agostino and O.~Luongo]{S.~Capozziello,$^{1,2,3}$\thanks{capozzie@na.infn.it} R.~D'Agostino$^{1,2}$\thanks{rocco.dagostino@na.infn.it} and O.~Luongo$^4$\thanks{orlando.luongo@lnf.infn.it} \\
$^1$ Dipartimento di Fisica “E. Pancini”, Università di Napoli “Federico II”, 80126 Napoli, Italy.\\
$^2$ INFN, Sezione di Napoli, Complesso Universitario di Monte S. Angelo, Via Cintia Edificio 6, 80126 Napoli, Italy.\\
$^3$Laboratory for Theoretical Cosmology,
Tomsk State University of Control Systems and Radioelectronics (TUSUR),
634050 Tomsk, Russia,\\
$^4$ INFN, Laboratori Nazionali di Frascati, 00044 Frascati, Via Enrico Fermi, Italy.
}

\date{Accepted XXX. Received YYY; in original form ZZZ}

\pubyear{2020}

\begin{document}

\label{firstpage}
\pagerange{\pageref{firstpage}--\pageref{lastpage}}
\maketitle

\begin{abstract}
Cosmography becomes non-predictive when cosmic data span beyond the red shift limit $z\simeq1 $. This leads to  a \emph{strong convergence issue} that jeopardizes its viability. In this work, we critically compare  the two main solutions of the convergence problem, i.e. the $y$-parametrizations of the redshift and the alternatives to Taylor expansions based on Pad\'e series. In particular, among several possibilities, we consider two widely adopted parametrizations, namely $y_1=1-a$ and $y_2=\arctan(a^{-1}-1)$, being $a$ the scale factor of the Universe. We find that the $y_2$-parametrization performs relatively better than the $y_1$-parametrization over the whole redshift domain. Even though $y_2$ overcomes the issues of $y_1$, we get that the most viable approximations of the luminosity distance $d_L(z)$ are given in terms of Pad\'e approximations. In order to check this result  by means of cosmic data, we analyze the Pad\'e approximations up to the fifth order, and  compare these series with the corresponding $y$-variables of the same orders. We investigate two distinct domains involving Monte Carlo analysis on the Pantheon Superovae Ia data, $H(z)$ and shift parameter measurements. We conclude that the (2,1) Pad\'e approximation is  statistically the optimal approach to explain low and high-redshift data, together with the fifth-order $y_2$-parametrization. At high redshifts, the (3,2) Pad\'e approximation cannot be fully excluded, while the (2,2) Pad\'e one is essentially ruled out. 
\end{abstract}

\begin{keywords}
{cosmological parameters - dark energy - observations}
\end{keywords}


\section{Introduction}
\label{sec:intro}

The Universe dynamics is currently undergoing an accelerated phase driven by a dark energy component, which can assume    the form of a cosmological constant ($\Lambda$) originated by the quantum fluctuations of  early vacuum \citep{Sahni00,Carroll01,Peebles03,Copeland06,luongobis,luongo}. A wide experimental evidence suggests the concordance \emph{flat} $\Lambda$CDM paradigm as the most successful model to explain early and late time dynamics. However, the recent developments of the Planck collaboration \citep{PlanckInfla} seems to indicate that  corrections to Einstein's gravity can explain the cosmological inflation immediately after the Big Bang and confirm the so called Starobinsky model \citep{starobinsky}. Hence, possibility that a slightly evolving dark energy contribution can fuel the energy momentum-tensor, alleviating the problems related to the today tiny value\footnote{This heals {\emph de facto} the thorny issue of fine-tuning plaguing the model \citep{Weinberg89}.} of $\Lambda$, is still valid. Thus, distinguishing dark energy from $\Lambda$ and understanding whether the equation of state of the Universe is varying \citep{dark3,dark4,dark2,dark5,dark1bis,dark1,dark6,dark1tris,dark6bis,piedipalumbo} have widely prompted the search for model-independent treatments and null-diagnostics \citep{alg1,alg4,alg1quatris,alg1tris,alg2tris,alg2,alg3,alg1bis,alg2bis,alg4bis,esc1,esc2,lobo}. Model-independent techniques are constructed to give hints on the correct form of dark energy. However, the main issue  is related to  the origin of such a dark component which could come from some material  field or be the cumulative effect of  some modified   gravity  \citep{curvature, nojiri, amendola, oikonomou}.

Among the possible model-independent strategies, great emphasis has been given to the well-consolidated approach named \emph{cosmography} \citep{Visser,LuongoReview, rocco_review}. The advantage of cosmography consists in assuming only the homogeneity and isotropy of the Universe\footnote{Cosmography assumes the validity of the cosmological principle and turns out to be purely model-independent in terms of derivatives of the scale factor.}, factorizing the scale factor around the present time.
However, two main problems are associated to the use  of  the above expansion and analysis of  the cosmic data. First, distinguishing the evolution of dark energy from $\Lambda$ requires highly refined limits over the derivatives of $a(t)$. So, one needs a wide number of data points to reduce cosmic systematics and to provide constraints over the cosmographic coefficients. In this sense, the lack of many and precise data limits the use of cosmography. A way of improving the quality of numerical fits may be using combined data sets or mock compilations based on the sensitivity and design of future surveys \citep{rocco_mock,rocco_mockbis}. The second problem, is based on the form of data. In fact, in order to determine possible departures from the standard cosmological model, one takes data which exceed the limit $t\simeq t_0$, with $t_0$ today epoch,  or, alternatively speaking, the bound $z\simeq0$. In other words, one faces a convergence problem caused by data that are far from the limits of the Taylor expansions, leading to a severe error propagation, which reduces the cosmographic predictions \citep{aert}. To solve this problem, several approaches have been  proposed so far in the literature. One of these relies  on the use of auxiliary variables. The idea is to re-parametrize the redshift variable through functions of $z$, and expand in series of the cosmological observables. These functions should vanish at $z=0$ and converge to a finite number at $z\rightarrow\infty$. Another relevant possibility is to consider a smooth evolution of the involved observables by expanding them in terms of rational approximations. A feasible consequence of this scheme is the stability of these new expansions over a large redshift interval. Examples of this approach are the Pad\'e and Chebyshev rational polynomials \citep{pade2,pade_luongo,chebyshev,pade3,Sen,rocco_review,Micol}.

In this paper, we want to show how to significantly reduce the convergence problem, overcoming the weakness of cosmography by constructing the most suitable expansions to fix refined bounds over cosmographic coefficients. Thus, to handle the convergence issue, we first theoretically investigate the advantages of rational polynomials against auxiliary variables. In particular, we show that Pad\'e polynomials well adapt to match the cosmic dynamics at high redshifts. Based on this result which substantially  agrees with previous ones  in  literature, we proceed to select the most suitable order of Pad\'e expansion. To do so, we compare the two classes of approaches, i.e. the one making use of the auxiliary variables and the second using Pad\'e series of different orders. We therefore propose a few conditions that every rational approximation should fulfill in order to be more predictive than a given parametrization. Moreover, we discuss that any new class of auxiliary variables, extending the role played by $y_1$ and $y_2$, turn out to be less predictive than Pad\'e approximations. In particular, we split our discussion considering  two regimes corresponding to low and high redshifts. In the low-redshift regime, we note that the results of the third-order $y_1$-variable are substantially different by $2\sigma$ from the corresponding Taylor constraints, whereas the (2,1) Pad\'e polynomial turns out to be the best-performing third-order approximation. We compare $y_2$ with $y_1$ and we find that, at the fourth-order expansion, still the $y_2$-variable behaves better than the $y_1$-variable, while a persistent disfavour against
the (2,2) Pad\'e polynomial is present, if compared to the (2,1) Pad\'e polynomial.
At the fifth-order expansion, all the techniques are characterized by large uncertainties on the parameters beyond $s_0$, leaving open the possibility to use (3,2) Pad\'e polynomial, but disfavouring the corresponding $y$-variables. At high redshifts, we find that the most suitable approach is the (2,1) Pad\'e polynomial, showing, in all cases, a very strong evidence against the $y$-variables. However, for the sake of clearness, a relevant fact is offered by the $y_2$ variable which, at the fifth order, seems to better frame the cosmographic curves. In any case, we rule out the use of the (2,2) Pad\'e rational polynomial that does not represent a suitable approximation of $d_L(z)$. We statistically analyze all the predictions of the different approximations by means of Bayesian selection criteria. In view of our theoretical considerations and the statistical inference results, we conclude that  cosmography built upon (2,1) Pad\'e polynomial remains the most suitable one at both low and high-redshift domains. To quantitatively show this, we involve the most recent Supernova Ia data, the direct $H(z)$ measurements and the early time measurements of the cosmic microwave background (CMB) shift parameter.

The paper is structured as follows. After this introduction, we develop the main features of the cosmographic technique in  \Cref{sec:sezione2}. In particular, we discuss how to build up the cosmographic series and the basic demands of cosmography. We then face the convergence problem of the cosmographic series in \Cref{sec:sezione3}, and we discuss the possible solutions to it. In particular, we introduce the concept of rational approximations and auxiliary variables, dealing with the concepts of Pad\'e and $y$-cosmography respectively. We then show how to construct theoretically the most suitable approximation, by means of Pad\'e polynomial first and then through the $y$-variables. In \Cref{sec:sezione4}, we develop our numerical analyses by means of a hierarchy among coefficients. We handle different orders, starting from the third one, and then increasing it up to the fifth order. For each order, we analyze separately low and high-redshift data and we find constraints and convergence for each approximations. Hence, we compare our expectations with the numerical results, and the different orders by means of statistical criteria. Finally, the conclusions and perspectives of our work are reported in \Cref{sec:sezione5}. In the Appendix, the cosmographic expansions, up to fifth order, are reported.

Throughout the paper, we use natural units with $c=1$.


\section{The cosmographic approach}
\label{sec:sezione2}

To describe the homogeneous and isotropic universe, we consider the Friedmann-Lema\^itre-Robertson-Walker (FLRW) line element, given by:
\begin{equation}
ds^2=dt^2-a(t)^2\left[\dfrac{dr^2}{1-kr^2}+r^2\,dl^2\right] ,
\end{equation}
where the present-day value of the scale factor is conventionally normalized to the unity (i.e. $a_0=1$), and $dl^2\equiv (d\theta^2+\sin^2\theta\ d\phi^2)$. Here, $k$ defines the spatial curvature. Dynamics of the universe  is determined by solving  the Friedmann equations:
\begin{equation}\label{eq:Friedmann}
H^2=\frac{1}{3}\rho-\frac{k}{a^2}\ ,\qquad\quad \dot H+H^2 = -\frac{1}{6}\left(3P+\rho\right)\,,
\end{equation}
where $H\equiv \dot{a}/a$ is the Hubble expansion rate; $\rho$ and $P$ are the energy density and pressure of the cosmic fluid, respectively.
The cosmological dynamical system is completed considering  the continuity equation $\dot{\rho}+3H(\rho+P)=0$ and  the equation of state $w_i(z)=P_i/\rho_i$, where the index $i$  represents the fluid species sourcing  the Universe. It is then convenient to define the density parameters $\Omega_i\equiv \rho_i/\rho_c$, normalized to the critical density $\rho_c\equiv 3H_0^2/(8\pi G)$.

According to the observed densities, the  content of the Universe can be assumed as made of dust matter and dark energy, while we can neglect the contribution of radiation at late times. Consistently with most recent cosmological observations \citep{Planck18}, we also assume that the universe has vanishing curvature $(k=0)$ also if, recently, there are claims that it could be $k\neq 0$ \citep{Melchiorri} but this result has to be definitely proved.
The standard strategy adopted in cosmography is to expand  the scale factor in Taylor series around the present time, namely the age of the Universe ($t_0$), to study the accelerated phase of the cosmic expansion\footnote{The cosmographic technique allows to study the evolution of the Universe in a model-independent way by means of kinematic variables \citep{Visser}. In other words, cosmography does not require the assumption of any specific cosmological model to investigate the behaviour of dark energy. In this picture, the equation of state of the cosmic fluid is not postulated \emph{a priori}, and the only assumption concerns the validity of the cosmological principle.}:
\begin{equation}
a^{(3)}= \left[ 1 + H_0 \; (t-t_0) - {1\over2} \; q_0 \; H_0^2 \;(t-t_0)^2+{1\over3!}\;  j_0\; H_0^3 \;(t-t_0)^3 \right]
\label{eq:scale factor jerk}
\end{equation}
From \Cref{eq:scale factor jerk}, which represents the third-order expansion of the scale factor, one can define the cosmographic series, i.e. the series of coefficients in terms of derivatives of the scale factor. These terms are defined from symmetry principles \citep{Weinberg}, without invoking specific solutions to Eqs. \eqref{eq:Friedmann}:
\begin{equation}\label{toto}
H(t)\equiv \dfrac{1}{a}\dfrac{da}{dt}, \hspace{0.5 cm} q(t)\equiv -\dfrac{1}{aH^2}\dfrac{d^2a}{dt^2},\hspace{0.5 cm} j(t) \equiv \dfrac{1}{aH^3}\dfrac{d^3a}{dt^3},
\end{equation}
These coefficients are known as \textit{Hubble}, \textit{deceleration} and \textit{jerk} parameters. In particular, from the sign of $q$, one can infer whether the universe is decelerating, i.e. $q_0>0$, or accelerating, i.e. $q<0$; a positive sign of $j$ indicates a transition time between the two phases. More recently, the cosmographic series has been extended in order to compare cosmological models that degenerate with respect to $q_0$ and $j_0$. In other words, one has the need to compute more than three cosmographic coefficients to show possible deviations from the standard cosmological model. So, extending \Cref{eq:scale factor jerk} up to the fifth order, we have
\begin{equation}
a^{(5)}= a^{(3)}+\left[{1\over4!}\;  s_0\; H_0^4 \;(t-t_0)^4+{1\over5!}\;  l_0\; H_0^5 \;(t-t_0)^5\right] ,
\label{eq:scale factor lerk}
\end{equation}
where we have introduced
\begin{equation}
s(t)\equiv \dfrac{1}{aH^4}\dfrac{d^4a}{dt^4}, \hspace{1cm} l(t)\equiv \dfrac{1}{aH^5}\dfrac{d^5a}{dt^5},
\label{toto2}
\end{equation}
colled the \emph{snap} and \emph{lerk} parameters, respectively. The possibility to discriminate between a given model and the standard cosmological one passes through taking into account at least $a^{(4)}$ according to the present state of observations.

\subsection{The standard approach to cosmography}

With the above considerations, one can use the definition of the redshift in terms of the scale factor, $z=a^{-1}-1$, to relate the Hubble parameter to the luminosity distance $d_L(z)$ as
\begin{equation}
H(z)=\Big[\frac{d}{dz}\left(\frac{d_L(z)}{1+z}\right)\Big]^{-1}\,,
\end{equation}
 to obtain
\begin{align}\label{dlandH}
d_L(z)&=\frac{z}{H_0}\sum_{n=0}^{N}{\alpha_n\over n!}z^n\,,\\
H(z)&=\sum_{m=0}^{M}{H^{(m)}\over m!}z^m\,,
\end{align}
or, explicitly, at the  $5^{th}$ order of expansion in $d_L(z)$, one obtains
 \begin{equation}
d_L^{(5)}=\frac{z}{H_0}\left(\alpha_0+\alpha_1z+\alpha_2\frac{z^2}{2}+\alpha_3\frac{z^3}{6}+\alpha_4\frac{z^4}{24}\right)\,,
\end{equation}
or, at the $4^{th}$ order of expansion in $H(z)$, we have\footnote{We consider $\mathcal E\equiv H/H_0$.}
\begin{equation}
\mathcal E^{(4)}\simeq 1+H_1z+H_2\dfrac{z^2}{2}+H_3\dfrac{z^3}{6}+H_4\dfrac{z^4}{24}\,.
\end{equation}
Therefore, by means of the definitions \eqref{toto}-\eqref{toto2}, one can write the coefficients $\alpha_i, H_i$ in terms of the cosmographic parameters (see \Cref{appendix} for details) and compare the above formulae directly with observations. The difference in the expansion order between $d_L$ and $H$ is due to the definition of $d_L$ which gives rise to  some  additional complexity with respect to fitting directly through $H$. For these reasons,  a given order of $d_L(z)$  provides much more statistical discrepancies with respect to $H(z)$ of the corresponding order \citep{altrocosmo,altrocosmobis,altrocosmotris}.

This fact implies errors which are produced by the  truncations and so they propagate as $\sim\mathcal O(6)$ and $\sim\mathcal O(5)$ for $d_L$ and $H$ respectively. In fact, systematics induced by truncating the series at a certain order may influence the cosmographic analysis, biasing the corresponding numerical results. If, from  one hand, the convergence decreases when additional higher-order terms are introduced, on the other hand, the accuracy of the analysis may be compromised by considering only lower orders. A possible solution to this issue is to analyze series at different orders and thus constrain different sets of parameters, considering some sort of \emph{hierarchy} among series. In the following, we will consider this approach and we will constrain different orders to check the statistical and numerical differences of the produced outcomes.

Let us conclude this section by discussing the role of spatial curvature in the cosmographic context. Degeneracy among coefficients and spatial curvature is due to the fact that all the cosmographic parameters are related to the Hubble rate. In particular, the contribution of the Hubble constant in the luminosity distance can be factorized into $d_L=d_H\times\tilde{d}_L(z; q_0,j_0,\hdots)$, where $d_H\equiv H_0^{-1}$. Then, $d_H$ becomes an additional  constant as one fits with respect to supernovae, requiring that a single data set is unable to break the degeneracy among coefficients. The same happens when one considers spatial curvature which degenerates with $j$ and further orders, and so one cannot bound them only without fixing spatial curvature. To heal this problem, in this work, we assume $k=0$, in agreement with the CMB observations which confine its value to a very small interval around zero \citep{Planck18}. The aim is  to compute single coefficients instead of combinations between cosmographic series and $k$. However, although this scheme is well consolidated and works fairly, we still do not have hints on how to perform a direct high-redshift cosmography. The former is intimately related to the \emph{problem of convergence}, consisting on the issue that affects the cosmographic series when data beyond the limit $z<1$ are considered. To better realize this, it is easy to notice that the Taylor series has a limited convergence radius. In the next section, we face the convergence problem and we discuss how to overcome it by extending the Taylor series up to high-redshift domains using strategies over the expansions themselves.

\section{Towards a solution of the cosmographic issues}
\label{sec:sezione3}

According to what discussed above, cosmographic Taylor series are expected to suffer from divergence problems as cosmic data exceed the limit $z<1$. In other words, truncating series that are obtained by expansions around $z=0$ causes difficulties to analyze data at $z\geq 1$. This fact leads to systematic error propagations and limits the numerical results which are affected by enhanced bad convergence.  The convergence issue is the main limitation of cosmography, and so it becomes of fundamental importance overcoming it by building up some  more reliable cosmographic treatment. To do so, one can employ rational polynomials or new parametrizations of the redshift variable\footnote{This is possible by means of \emph{auxiliary variables}, which maps the domain $z\in(0,\infty)$ into a sphere of convergence radius $R_\rho\leq1$.}, with the final aim of extending the convergence radii of standard Taylor expansions. Both rational approximations and parametrizations of $z$ are artificial and need specific techniques to be taken into account.

We now show that, albeit widely used to overcome the convergence problem, auxiliary variables are much less predictive than particular classes of rational approximations. In particular, we confront both the approaches showing that auxiliary variables are not adequate to deal with cosmological data at high redshifts, although they have been introduced with this purpose.

\subsection{First solution: rational polynomials}

Making cosmographic expansions stable at high redshifts can be possible as one assumes $(n,m)$ rational approximations of a given cosmological observable. Indeed, approximations are built from the ratio between a $n$-th and a $m$-th order polynomials, leading to a overall order, $n+m$, which univocally defines the set of parameters entering cosmological fits. The construction of rational approximations is jeopardized by the issue of degeneracy among the cosmographic coefficients. To picture this fact, let us take as an example ${d_L}_{(m=1)}^{(n=2)}$. In terms of Taylor expansions, it leads to a third-order approximation defined by three cosmographic parameters. However, there exists a degeneration between the orders, as also ${d_L}_{(m=2)}^{(n=1)}$ is equivalent to its Taylor analogue, albeit ${d_L}_{(m=1)}^{(n=2)}$ and ${d_L}_{(m=2)}^{(n=1)}$ are different between them.

This fact becomes crucial when high redshift data are involved. In fact, for small redshifts, rational expressions are essentially indistinguishable to the Taylor one, but, as high-redshift surveys are considered, the situation dramatically changes. Here we propose a first criterion that one should consider to choose the correct rational approximation. We need the convergence radius of rational polynomials to be equivalent or higher than the corresponding Taylor series, determining the following practical rule:
\begin{itemize}
  \item \emph{The most suitable rational approximation corresponds to the function that maximizes the convergence radius, once the orders $(n,m)$ are chosen.}
\end{itemize}
The advantage is easy to understand: the stability of the approximation is intimately related to its convergence radius, and thus stable rational approximations would give stable fitting procedures.
In principle, however, this is valid for an infinite number of parameters only, namely as the original function is fully-recovered. As a consequence, this issue provides a second practical rule:
\begin{itemize}
  \item \emph{The most suitable rational approximation is the one that minimizes the number of parameters of $m$, taking into account the first condition mentioned above.}
\end{itemize}
In fact, the more parameters are  in the denominator, the larger is the freedom (degeneration) in the choice of intervals that can nullify the denominator, creating poles. When the number of parameters is small, one can better handle this situation. In other words, the denominator, i.e. the only part of rational approximations which can create poles, might be stable. Thus, a small number of free parameters is essential to enable the rational approximation to be convergent. The two aforementioned conditions translate in finding the \emph{best compromise between arbitrary-order expansions and minimal number of free parameters in the denominator}.

Likely the simplest choice is then to consider the (2,1) rational polynomial characterized by only one parameter in the denominator, i.e. minimizing $m$ as the overall order is fixed. However, this provides additional issues in the choice of the best polynomials. In fact, other choices, e.g. (3,1) and (4,1) polynomials, are equally valid, in principle, and in the following we ought to explain why the (2,1) polynomials should be preferred over the (3,2) or the (1,1) and (2,2) and so forth from a statistical point of view.

\subsubsection{The Pad\'e approximation}

As a relevant example, we present the method of the Pad{\'e} approximations \citep{Baker96}. The Pad\'e technique is built up from the standard Taylor definition and is used to lower divergences at $z\geq1$. Thus, a given function $f(z)=\sum_{i=0}^\infty c_iz^i$, expanded with a given set of coefficients, namely $c_i$, is approximated by means of a $(n,m)$  Pad{\'e} approximant by the ratio
\begin{equation}
P_{n, m}(z)=\dfrac{\displaystyle{\sum_{i=0}^{n}a_{i} z^{i}}}{1+\displaystyle{\sum_{j=1}^{m}b_j z^{j}}}\,,
\label{eq:def_Padé}
\end{equation}
where the Taylor expansion matches the coefficients of the above expansion up to the highest possible order:
\begin{align}\label{am}
&P_{n,m}(0)=f(0)\,,\\
&P_{n,m}'(0)=f'(0)\,,\\
&\vdots	\\	
&P_{n,m}^{(n+m)}(0)=f^{(n+m)}(0)\,.
\end{align}
By construction, in the numerator we have $n+1$ independent coefficients, whereas in the denominator $m$, for a total of  $n+m+1$ unknown terms. From
\begin{equation}
\sum_{i=0}^\infty c_iz^i=\dfrac{\displaystyle{\sum_{i=0}^{n}a_{i} z^{i}}}{1+\displaystyle{\sum_{j=1}^{m}b_j z^{j}}}+\mathcal{O}(z^{n+m+1})\,,
\end{equation}
and, then,
\begin{equation}
(1+b_1z+\ldots +b_mz^m)(c_0+c_1z+\ldots)= a_0+a_1z+\ldots+a_nz^n +\mathcal{O}(z^{n+m+1})\ ,
\label{coeff}
\end{equation}
equating the same power coefficients, one gets $n+m+1$ equations with $n+m+1$ free variables. The advantages of Pad\'e rational approximations are thus:

\begin{itemize}
  \item \emph{the polynomials approximate flexes and singular points in a better way than Taylor;}
  \item \emph{the polynomials reduce error bias propagating as data are outside the limit $z<1$;}
  \item \emph{the polynomials, independently from choosing appropriate orders, hold for all terms to fit in the numerator.}
\end{itemize}

Hence, Pad\'e polynomials are expected to extend the convergence radius of the Taylor series.
Another relevant fact is that Pad\'e series, among all rational approximations, address in a simple way how to choose the orders $(n,m)$. In fact, we note that in cosmography we approximate a generic function $f$ as $f\sim f_0+f_1+f_2+f_3+\ldots$, where $\left |f_3=f_3(q_0,j_0)\right | <\left | f_2(q_0)\right |$, due to the larger weight of the lowest orders with respect to the higher orders reflecting the lower stability of the less numerous high-redshift data. This implies that a rational function of the type $x/(1+x+x^2)$ approximatively behaves as $x/(1+x)$, or
$(x+x^2)/(1+x)\sim x+x^2$ and so on, with $x$ an arbitrary analytical variable. Numerator and denominator of the same order cause induced errors in the analysis as, for $x\rightarrow\infty$,
$N/D\rightarrow const$, where $N=N(x)$ and $M=M(x)$ are the polynomials in the numerator and denominator, respectively. This is the reason why \emph{rational approximations with the same order in the numerator and denominator do not provide accurate cosmographic results.}

Understanding why the (2,1) polynomial is preferred over the (3,2) and (3,1) orders is related to the degeneracy among coefficients.
By construction, in the Pad\'e series,
 all the free parameters are present already in the numerator.
Pad\'e series are of lower order than the corresponding Taylor polynomials. As an example, we mention that for (3,1) and (2,1) polynomials, one has respectively
\begin{equation}\label{31}
P_{3,1}(z)=\frac{a_0+a_1 z+a_2 z^2 + a_3 z^3}{1+b_1 z}\ ,
\end{equation}
and
\begin{equation}\label{21}
P_{2,1}(z)=\frac{a_0 +a_1z+a_2 z^2}{1+b_1 z}\ .
\end{equation}
In the (3,1) case, we have $|b_1|\ll |a_1|$ and $|b_1| \ll |a_2|$, while in the (2,1) case, we have $|a_1|\sim|a_2|\cdot|b_1|$.
As a result, in the (3,1) case,  the denominator does not enter the problem, whereas it does play a role in the (2,1) case.
Therefore, at low redshifts,  the numerical fits are influenced by all the parameters in the (2,1) case, contrary to the (3,1) case. This provides us with the following rule:
\begin{itemize}
  \item \emph{The rational approximation is, for an arbitrary choice of $z$ (let us say $z=1$ for simplicity), a Pad\'e series in which the ratio maximizes the condition $\frac{N}{D}\Big|_{z=1}\approx 1$, and is not $\gg 1$.}
\end{itemize}
The above condition is clearly verified for any order, also for the (2,1) case. This implies that
\begin{itemize}
  \item \emph{error bars increase as low-redshift data only are involved;}
  \item \emph{error bars decrease as high-redshift data are included.}
\end{itemize}
In fact, adding high-redshift data makes the denominator count more than the numerator, thus minimizing the errors with respect to the order. We can then summarize our findings as follows:

\begin{itemize}
  \item \emph{The Pad\'e series are suitable tools to address high-redshift cosmography}, since they involve rational polynomials which account for the aforementioned conditions.
  \item \emph{The most stable order of the Pad\'e series is (2,1)}; we can demonstrate this statement  based on the constructions of Pad\'e series and on mathematical rules derived from the degeneracy among coefficients.
  \item \emph{The Pad\'e series are characterized by larger uncertainties compared to the corresponding Taylor series.} This is evident as much as small data are involved into the problem, while becomes non-relevant as high-redshift data are involved. Unfortunately sets of cosmic data surveys with $z>1$ are typically poorly constrained, leaving cosmography non-predictive if only high-redshift data are included into experimental analyses.
  \item \emph{The errors associated to the Pad\'e series reduce as soon as  high-redshift data are included into the analysis.} The explanation derives from  the previous point: in the case of high-redshift data only, if the data set is large, Pad\'e series become stable and permit a high-redshift cosmography at arbitrary redshift.
\end{itemize}

To summarize, the problem of constructing a model-independent cosmography consists in adopting a rational approximant built up in terms of Pad\'e polynomial, which fulfills all the aforementioned requirements. In particular, the Pad\'e approximation that guarantees the best outcomes in the fitting procedures is given by the (2,1) polynomial. This leads to the problem that no more than $s_0$ can be accurately fitted even with higher redshift data, in agreement with previous works \citep{visser}. The question left open is therefore: does it exist an alternative strategy toward the determination of better convergent cosmographic series? Let us discuss this point by using  auxiliary variables in the next section.

\subsection{Second solution: auxiliary variables}

Another approach to heal the convergence problem is represented by the use of auxiliary $y$-variables. To build up a generic parametrization of $y$, we can write
\begin{equation}\label{uu}
y=\mathcal F(z)\ ,
\end{equation}
satisfying the conditions
\begin{align}
\mathcal F(\infty)&<\infty\,,\label{conditio12}\\
\mathcal F(0)&=\mathcal C<\infty\label{conditio12bis}\,.
\end{align}
The properties \eqref{conditio12}-\eqref{conditio12bis} are fundamental to construct a $y$-variable that reduces the convergence radius when $z\rightarrow\infty$.
The procedure to reformulate the luminosity distance, $d_L(y)$, in terms of \Cref{uu} is possible in two ways. The first, after Taylor-expanding around $z=0$ the function $d_L(z)$, consists in substituting $z$ in terms of $y$ to get $\tilde d_L(\mathcal F^{-1}(y))$. Then, expanding in series around $y(0)$ permits to get $d_L(y)$. This implies that the most suitable choice of $\mathcal C$ is $\mathcal C=0$, fulfilling the fact that our cosmic era is characterized by $z=\mathcal F=0$. The second procedure takes into account an exact version of $d_L(y)$ \emph{without expanding it}, as before,  around $z=0$. Only after this assumption, one can expand the exact luminosity distance, obtaining the same result of the first method. The equivalence of the two approaches is due to the analyticity of $d_L(z)$ and justifies why, at $z=0$, we should have $y=0$. From a theoretical point of view, \emph{the auxiliary variables are unable to reproduce the $\Lambda$CDM predictions}, being unable to be suitable alternatives to Taylor expansions in general. To see this fact, it is enough to notice that when $z\rightarrow\infty$, we have $\mathcal F<\infty$. So, the luminosity distance converges to a plateau value which departs of several orders of magnitudes, depending on the choice of $y$, from observations. To overcome this fact, an immediate example is offered by powers of $y$, say $y^n$ with $0<n<1$, that clearly converge less rapidly than $y$, but leave the caveat toward the understanding of the most suitable order $n$, introducing a new problem: inferring the best $n$ from observations.

So that, in principle, one can imagine to re-parametrize the redshift with a function that abolishes the second condition displayed in \Cref{conditio12bis}. An example has been considered in \citet{ryo,ryo2}, where the authors introduced a log-polynomial $y$-variable. In such a way, they expanded the series around $\log_{10}(1+z)$, instead of $z$. Although intriguing, this only leads to re-parametrizing $z$, instead of a robust and well-motivated approach against the convergence problem, leaving a cautionary tale for cosmographic applications at $z\rightarrow\infty$. In other words, if one does not take into account the second condition,  \Cref{conditio12bis}, it is not possible to alleviate in any cases the convergence issue.
At small and intermediate redshifts, the use of auxiliary variables is impracticable since errors increase dramatically, with no substantial improvements on the cosmographic coefficients that are still far from being predictive in disentangling dark energy from a pure cosmological constant.

Considering the standard approach of auxiliary variables, the basic assumptions to construct any pictures of $y$-cosmography are: the auxiliary variable must be one-to-one invertible when passing from the redshift to it; it should not exhibit any divergence feature for any value of $z$; any new parametrization should be smooth along the Universe evolution and no critical points have to appear as the Universe expands. Relevant examples of this approach are given by \citet{Cattoen07,Aviles12}. They are
\begin{subequations}\label{subeqs}
\begin{align}
y_1&=1-a\,,\\
y_2&=\arctan(1-a^{-1})\,.
\end{align}
\end{subequations}
When these choices are plugged into $d_L$, without a series expansion, one typically gets sources of errors. For example, avoiding the series expansion after considering the change $z=\mathcal F^{-1}(y)$ in $d_L$ produces a rational function that cannot be better than parametrizing the redshift by means of Pad\'e approximations. This is why the formers behave better, i.e. they can be interpreted as convergent classes of auxiliary variables.
This argument can be used to explain why $y=\arctan(z)$ represents a better parametrization than $y_1$ and than any powers of $y_1$. In fact, this function is not a rational polynomial of $z$ built up \emph{ad hoc}, but a real function, as well as the log-parametrization that has the advantage to converge at very large redshifts. Even in this case the convergence creates a plateau at $z\rightarrow\infty$ and so does not explain very well the Universe expansion history at all times, but only at small and intermediate redshifts. In general,  we can conclude with the following statement:

\begin{itemize}
\item \emph{The most suitable parametrization is the one that most closely reproduces $z$ at any intervals of the Universe evolution.}
\end{itemize}

Moreover, by means of similar arguments, to contrive a viable redshift parametrization, the following conditions must be satisfied:
\begin{enumerate}
 \item \emph{The luminosity distance curve should not behave too steeply in the interval $z<1$.}
 \item \emph{The luminosity distance curve should not exhibit sudden flexes.}
 \item \emph{The curve should be one-to-one invertible.}
\end{enumerate}
The last three requirements are fulfilled for both $y_1$ and $y_2$. The second auxiliary variable, $y_2$, works better than the previous one.
We are now ready to compare $y$-cosmography with Pad\'e approximations directly with cosmic data.

\section{Observational constraints and the Bayesian inference}
\label{sec:sezione4}

To analyze the stability and convergence of the different cosmographic techniques, let us  consider the approximations of the luminosity distance based on\footnote{We refer the readers to \Cref{appendix} for the explicit expressions of the luminosity distance and the Hubble expansion rate corresponding to the different cosmographic techniques.}:
\begin{figure*}
\centering
\subfigure[Pad\'e of orders $n+m=(2;3)$\label{subfig21}]{\includegraphics[width=3in]{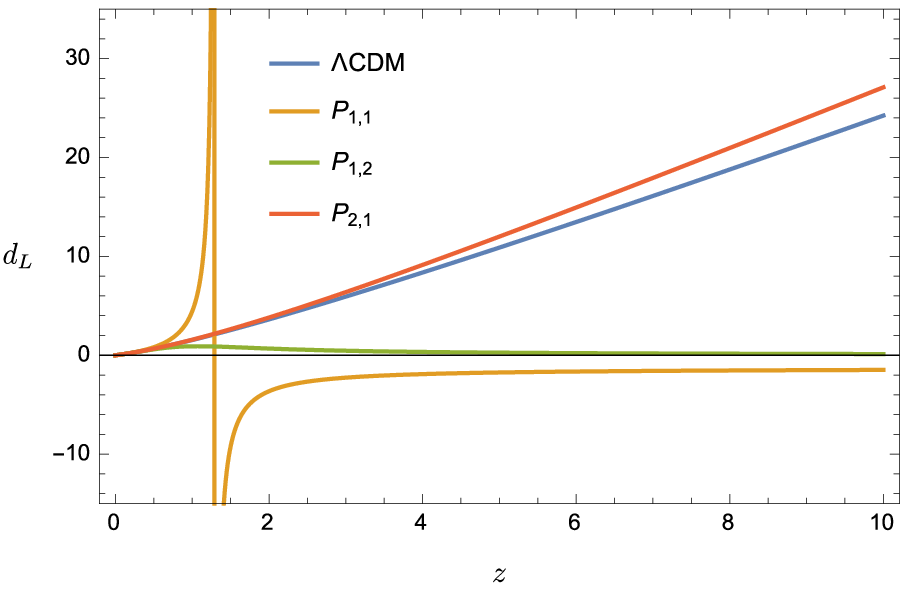}}
\subfigure[Pad\'e of orders $n+m=4$\label{subfig22}]{\includegraphics[width=3in]{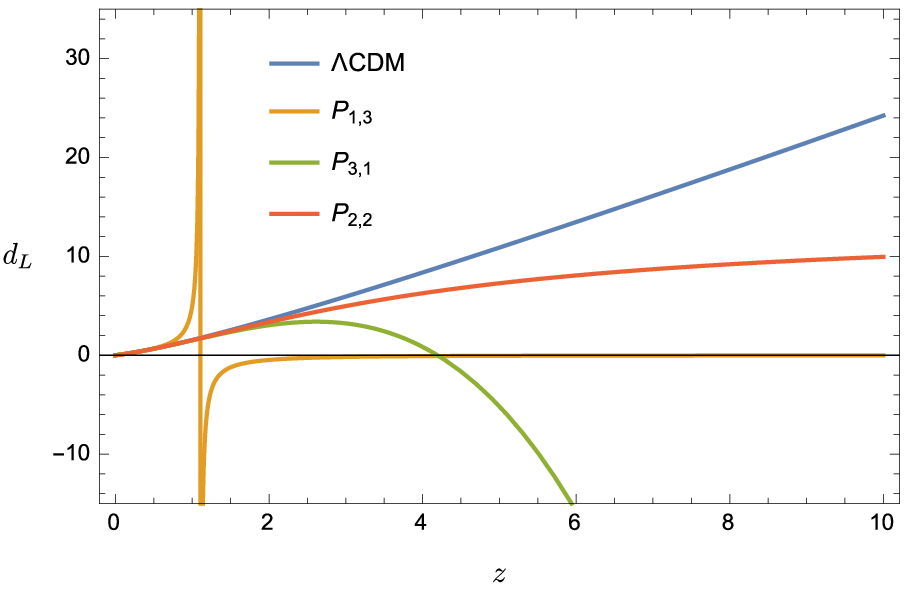}}
\subfigure[Pad\'e of orders $n+m=5$\label{subfig22}]{\includegraphics[width=3in]{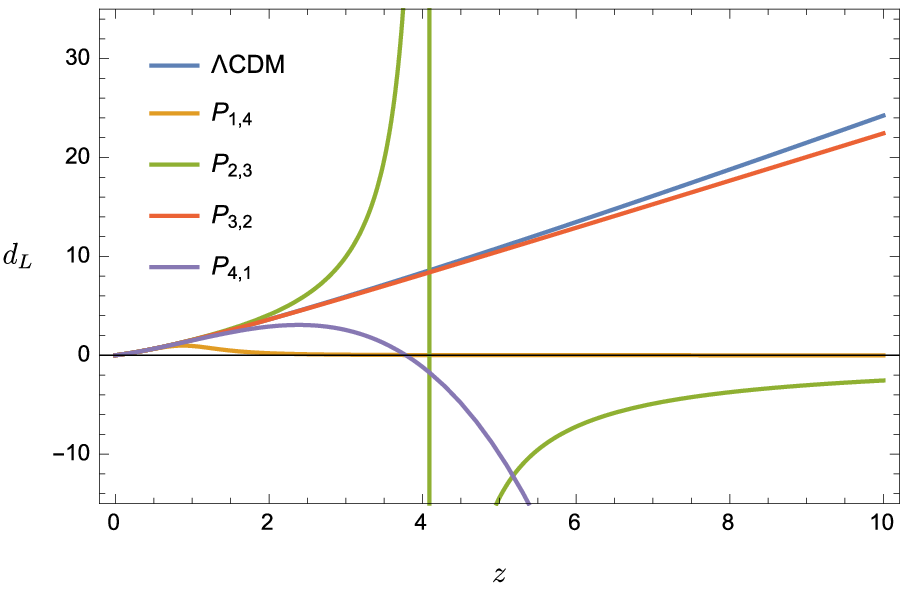}}
\caption{Redshift evolution of the Pad\'e approximations of the luminosity distance (in units of $H_0^{-1}$) up to the fifth order, using the fiducial set (\ref{eq:fiducial}). The prediction of the flat $\Lambda$CDM model is shown for comparison.}
\label{fig:Pade_curves}
\end{figure*}

\begin{itemize}
\item Taylor expansion of the third order $(T_3)$, fourth order $(T_4)$ and fifth order ($T_5$);
\item $y$-redshift expansions of the third order ($y_i^{(3)}$), fourth order ($y_i^{(4)}$) and fifth order ($y_i^{(5)}$) for the variables $y_1$ and $y_2$;
\item rational Pad\'e polynomials of the third order $(P_{2,1})$, fourth-order $(P_{2,2})$ and fifth order $(P_{3,2})$. These specific polynomials have been chosen according to their good behaviours over a large  interval of redshift (cf. \Cref{fig:Pade_curves}).
\end{itemize}

As a guideline,  we use the predictions of the flat $\Lambda$CDM model, characterized by the following Hubble expansion rate:
\begin{equation}
H_{\Lambda\text{CDM}}(z)=H_0\sqrt{\Omega_{r0}(1+z)^4+\Omega_{m0}(1+z)^3+1-\Omega_{r0}-\Omega_{m0}}\ ,
\label{eq:H_LCDM}
\end{equation}
where $\Omega_{r0}\approx 5\times 10^{-5}$ is the radiation density and $\Omega_{m0}$ is the present matter density parameter.

\noindent From the definitions (\ref{toto})-(\ref{toto2}) and \Cref{eq:H_LCDM}, it is possible to relate the cosmographic parameters to the physical quantity $\Omega_{m0}$ as
\begin{align}
&q_0 =  -1 + \frac{3}{2}\Omega_{m0}\ , \label{eq:derived_q0} \\
&j_0  =  1 \ ,  \label{eq:derived_j0} \\
&s_0 =  1 - \frac{9}{2} \Omega_{m0}\ ,  \label{eq:derived_s0} \\
& l_0=1 + 3\Omega_{m0} - \dfrac{27}{2} \Omega_{m0}^2\ .  \label{eq:derived_l0}
\end{align}
Moreover, assuming the concordance value $\Omega_{m0}=0.3$, one obtains the ``fiducial set'':
\begin{equation}
(q_0,\,j_0,\,s_0,\,l_0)=(-0.55,\,1,\,-0.35,\,0.685)\ .
\label{eq:fiducial}
\end{equation}

\noindent We therefore split our analysis in two stages, i.e. low and high-redshift domains.

\subsection{Low-redshift domain}

First, we considered the low-redshift regime through the use of the Pantheon type Ia Supernovae (SNe Ia) catalogue \citep{Scolnic18} and the observational Hubble data (OHD) acquired by means of the differential age method \citep{Jimenez02}. We refer the readers to \citet{D'Agostino19} and references therein for details on both SNe Ia and OHD data sets and their respective likelihood functions.
We thus performed a Monte Carlo Markov Chain (MCMC) integration on the combined SNe Ia+OHD likelihood.
\Cref{tab:low z} summarizes the numerical results for the cosmographic parameters up to the 95\% confidence level (C.L.).
It is interesting to compare these values with the values derived from \Cref{eq:derived_q0,eq:derived_j0,eq:derived_s0,eq:derived_l0} after fitting the $\Lambda$CDM model to the data. In particular, for the SNe Ia+OHD analysis, we find $\Omega_{m0}=0.295^{+0.026(0.058)}_{-0.029(0.051)}$, leading to the derived values reported in  \Cref{tab:low z} for the $\Lambda$CDM model. Moreover, in \Cref{fig:low z}, we show the $1\sigma$ and $2\sigma$ contours in the 2-D parameter space divided by orders of expansion.

\begin{table*}
\begin{center}
\footnotesize
\setlength{\tabcolsep}{0.5em}
\renewcommand{\arraystretch}{2}
\begin{tabular}{c c c c c c c c}
\hline
\hline
Model & $H_0$ &  $q_0$  &  $j_0$ & $s_0$ & $l_0 $ & $\Delta$AIC & $\Delta$BIC \\
\hline
$\Lambda$CDM & $69.2^{+1.9(3.8)}_{-1.9(3.8)}$ & $-0.56^{+0.04(0.09)}_{-0.04(0.08)}$  & 1 & $-0.33^{+0.13(0.26)}_{-0.12(0.23)}$ & $0.71^{+0.13(0.29)}_{-0.14(0.25)}$ & 0 & 0 \\
\hline
$T_3$ & $69.2^{+2.0(3.9)}_{-2.0(3.8)}$& $-0.58^{+0.08(0.16)}_{-0.08(0.15)}$   &  $1.02^{+0.24(0.49)}_{-0.24(0.46)} $  & - & - & 1.28 & 2.89\\
$T_4$ &  $69.3^{+2.0(3.9)}_{-2.0(3.9)}$ & $-0.66^{+0.13(0.26)}_{-0.13(0.25)}$  & $1.58^{+0.83(1.63)}_{-0.83(1.58)} $ & $0.97^{+1.15(3.54)}_{-1.88(2.76)}$ & - & 2.47 & 5.69\\
$T_5$ & $69.3^{+2.0(3.9)}_{-2.0(3.8)}$ & $-0.58^{+0.10(0.21)}_{-0.10(0.20)}$  & $0.98^{+0.54(0.88)}_{-0.48(0.99)}$ & $-0.63^{+1.49(2.67)}_{-1.57(2.64)}$ & $4.94^{+2.53(4.27)}_{-2.12(4.75)}$ & 2.98 & 7.81 \\
\hline
$y_1^{(3)}$ & $69.8^{+2.0(4.0)}_{-2.0(3.9)} $ & $-1.10^{+0.16(0.30)}_{-0.16(0.30)}  $ & $8.08^{+1.49(2.92)}_{-1.50(2.76)}$& - & - & 16.5 & 18.1 \\
$y_1^{(4)}$ & $69.6^{+2.1(4.2)}_{-2.1(4.1)} $ &  $-0.30^{+0.11(0.21)}_{-0.11(0.21)} $ & $0.26^{+1.39(2.39)}_{-1.17(2.58)}$ &  $7.49^{+4.63(8.33)}_{-4.83(8.17)}$ & - & 19.4 & 22.6 \\
$y_1^{(5)}$ & $69.5^{+2.0(3.9)}_{-2.0(3.8)} $ & $-0.75^{+0.11(0.23)}_{-0.11(0.21)}$ & $2.30^{+0.90(1.68)}_{-0.87(1.82)} $ & $0.21^{+4.47(7.80)}_{-4.07(7.72)}$ & $0.64^{+2.65(4.86)}_{-2.32(5.27)}$ & 4.59 &  9.42 \\
\hline
$y_2^{(3)}$ & $69.7^{+2.0(3.9)}_{-2.0(3.7)}$ & $-0.81^{+0.10(0.20)}_{-0.10(0.20)} $ & $2.82^{+0.49(0.99)}_{-0.49(0.89)}  $ & - & - & 7.71 & 9.32 \\
$y_2^{(4)}$ & $69.4^{+1.9(3.8)}_{-1.9(3.8)} $ & $-0.59^{+0.17(0.33)}_{-0.17(0.32)} $ & $0.56^{+1.59(2.72)}_{-1.59(2.69)}$ & $-3.58^{+2.11(5.83)}_{-3.13(4.80)}$ & - & 1.44 & 4.66  \\
$y_2^{(5)}$ & $69.2^{+2.0(4.0)}_{-2.0(3.9)} $ & $-0.55^{+0.14(0.31)}_{-0.16(0.29)} $ &  $0.54^{+0.96(1.72)}_{-0.78(1.91)} $ & $3.46^{+0.83(2.69)}_{-1.25(2.19)} $ & $4.64^{+2.29(4.62)}_{-2.47(4.55)}$ & 3.44 & 8.27 \\
\hline
$P_{2,1}$ & $69.3^{+2.0(3.9)}_{-2.0(3.8)} $ & $-0.73^{+0.13(0.26)}_{-0.13(0.26)} $ & $2.84^{+1.00(2.28)}_{-1.23(2.09)}$ & - & - & 1.20 &  2.81 \\
$P_{2,2}$ & $69.1^{+2.0(3.9)}_{-2.0(3.8)} $ & $-0.60^{+0.10(0.21)}_{-0.10(0.19)} $ & $1.53^{+0.81(1.59)}_{-0.81(1.57)}$ & $4.15^{+3.36(9.26)}_{-5.29(7.52)}$ & - & 3.17& 6.39  \\
$P_{3,2}$ & $69.1^{+1.9(3.8)}_{-1.9(3.8)} $ & $-0.60^{+0.10(0.19)}_{-0.10(0.19)}$ & $1.32^{+0.57(1.27)}_{-0.63(1.21)}$  & $8.47^{+1.52(4.56)}_{-2.28(3.80)}$ & $-2.1^{+7.7(11.8)}_{-4.3(16.4)}$ & 2.59 & 7.42 \\
\hline
\hline
\end{tabular}
\caption{MCMC results at the 68\% (95\%) C.L. for different cosmographic techniques from the low-redshift (SNe Ia+OHD) probes, compared to the derived predictions of the flat $\Lambda$CDM model. $H_0$ values are expressed in km/s/Mpc. The AIC and BIC values are computed with respect to the reference $\Lambda$CDM model.}
\label{tab:low z}
\end{center}
\end{table*}

\subsection{High-redshift domain}

\begin{figure*}
\centering
\subfigure[Third-order analyses\label{subfig25}]{\includegraphics[width=3in]{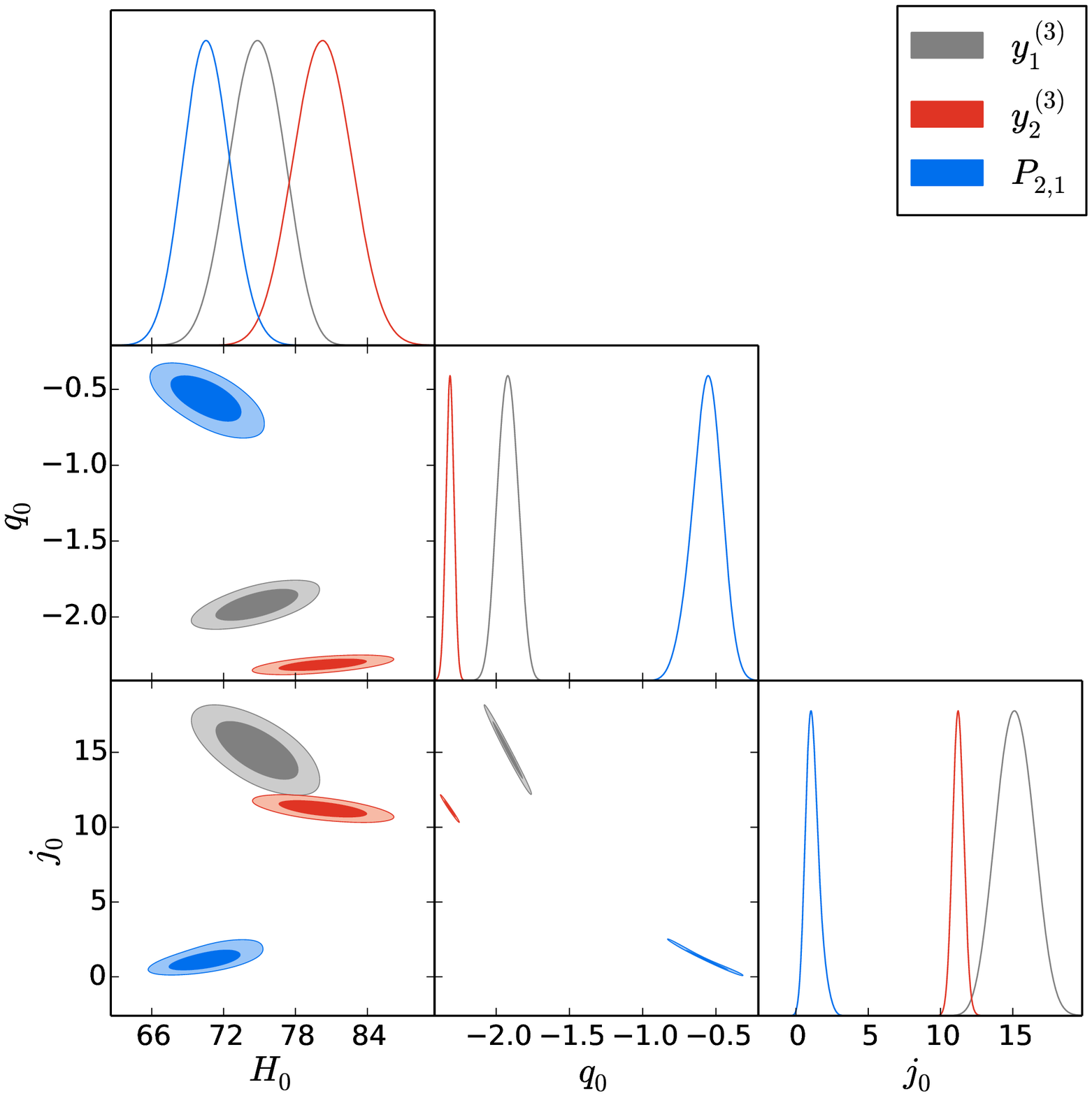}}
\subfigure[Fourth-order analyses\label{subfig25}]{\includegraphics[width=3.8in]{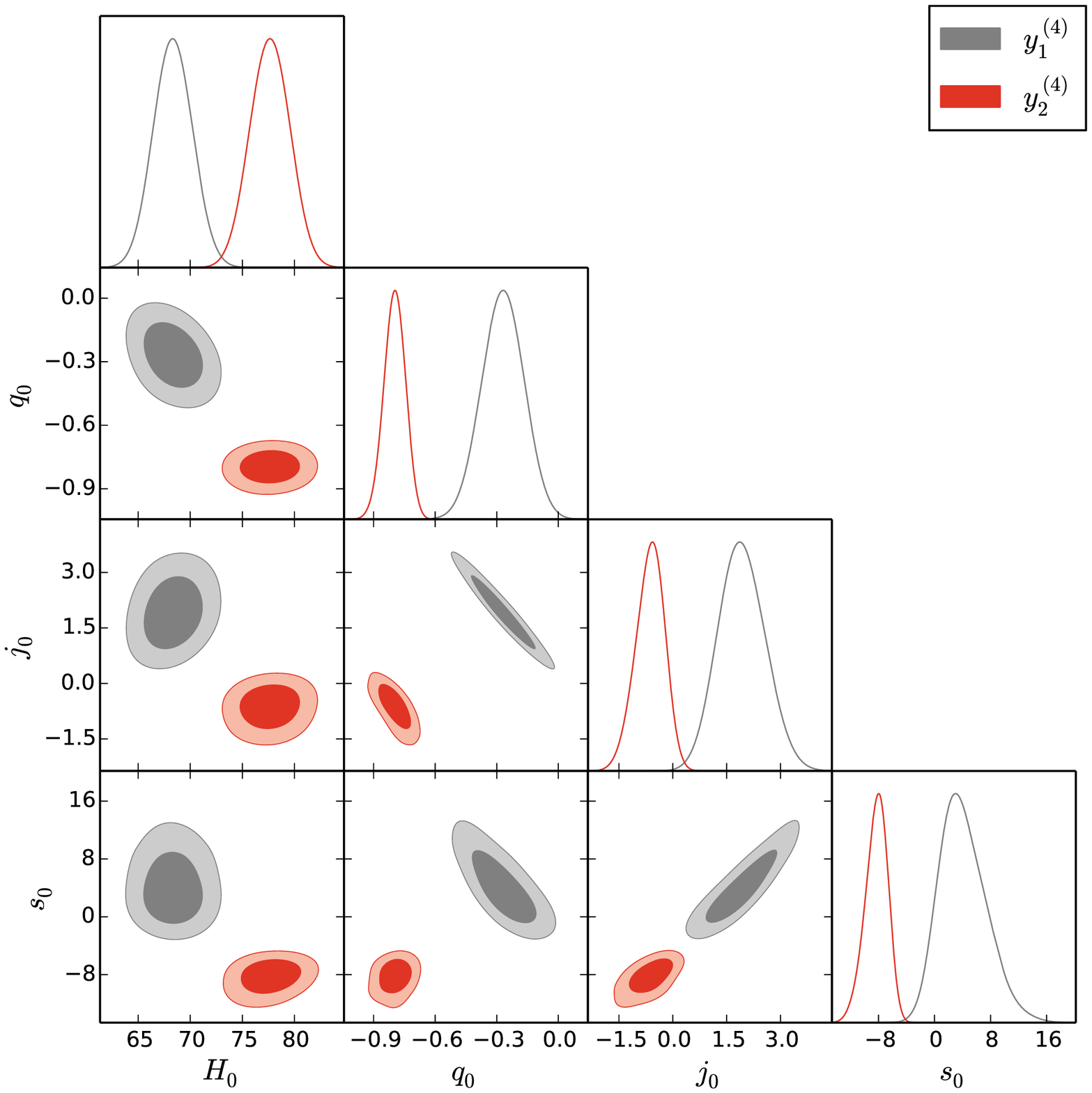}}
\subfigure[Fifth-order analyses\label{subfig25}]{\includegraphics[width=4.2in]{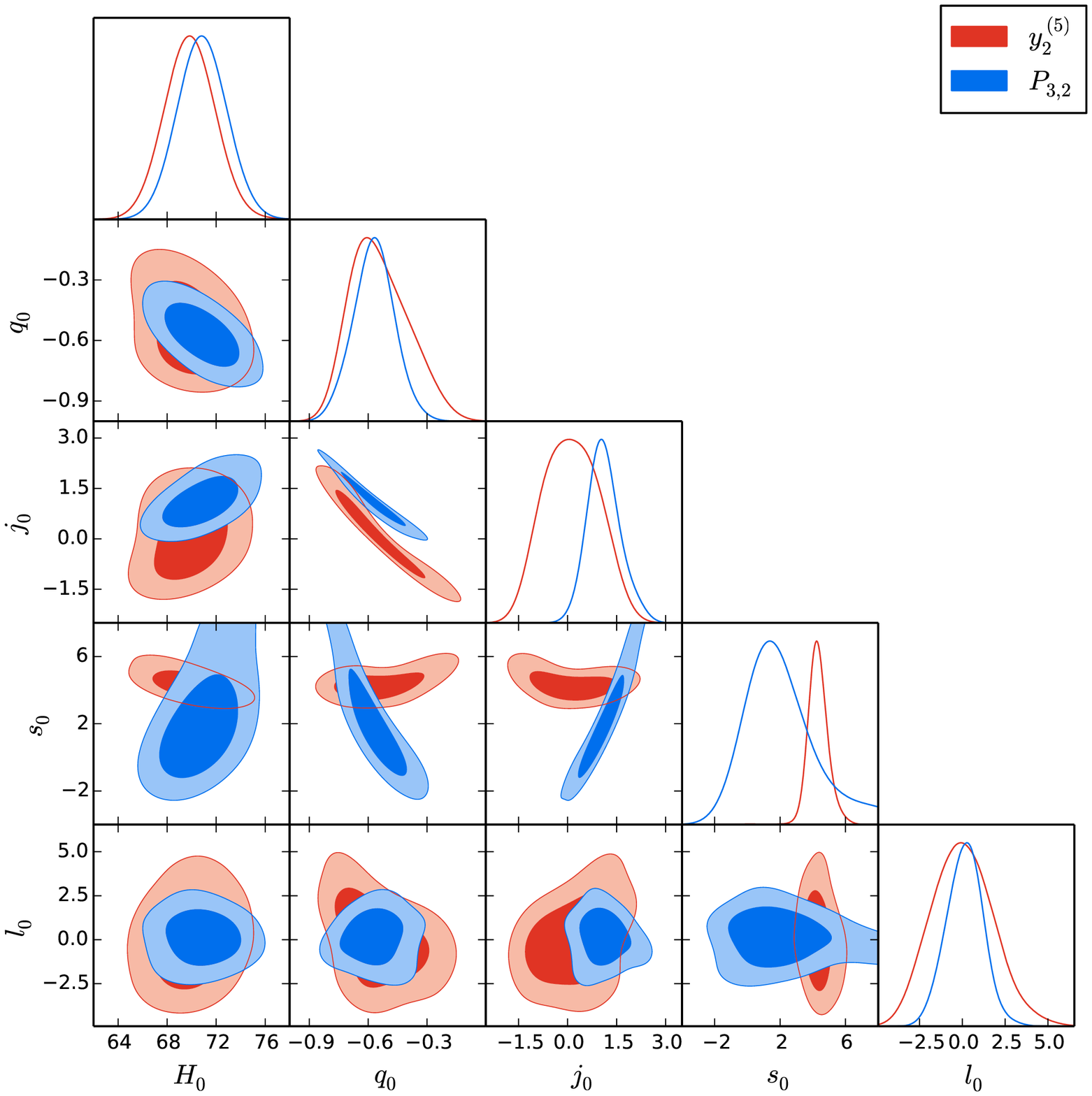}}
\caption{Combined low and high-redshift (SNe Ia+OHD+$\mathcal R$) constraints on the cosmographic parameters at the 68\% and 95\% C.L. for different techniques.}
\label{fig:high z}
\end{figure*}

To check also the high-redshift behaviour of our approximations on $d_L(z)$, we then take into account the CMB measurements by means of the shift parameter \citep{Efstathiou99}:
\begin{equation}
\mathcal R\equiv H_0\sqrt{\Omega_{m0}}\dfrac{d_L(z_{rec})}{(1+z_{rec})}\ ,
\end{equation}
which involves the luminosity distance at the epoch of recombination $(z_{rec})$.  The parameter $\mathcal R$  is effectively model-independent and insensitive to perturbations \citep{Maartens06}, and in our analysis we used the estimates of the Planck collaboration \citep{Planck15}, namely $\mathcal R=1.7488 \pm 0.0074$, together with $z_{rec}=1090.09$ and $\Omega_{m0}=0.315$. The results emerging from the combination of low and high-redshift data (SNe Ia+OHD+$\mathcal R$) are summarized in \Cref{tab:low+high z} up to the 95\% C.L., while \Cref{fig:high z} shows the $1\sigma$-$2\sigma$ contours in the 2-D parameter space. In this case, the reference $\Lambda$CDM values of the cosmographic parameters are derived from the measured value $\Omega_{m0}= 0.317^{+0.003(0.007)}_{-0.003(0.007)}$.

\begin{table*}
\begin{center}
\footnotesize
\setlength{\tabcolsep}{0.3em}
\renewcommand{\arraystretch}{2}
\begin{tabular}{c c c c c c c c}
\hline
\hline
Model & $H_0$ &  $q_0$  &  $j_0$ & $s_0$ & $l_0 $ & $\Delta$AIC & $\Delta$BIC \\
\hline
$\Lambda$CDM & $68.2^{+1.5(2.9)}_{-1.5(2.9)}$ & $-0.525^{+0.005(0.010)}_{-0.005(0.010)}$ & 1 & $-0.425^{+0.015(0.030)}_{-0.015(0.030)}$& $0.596^{+0.018(0.037)}_{-0.018(0.036)}$& 0 & 0 \\
\hline
$y_1^{(3)}$ & $74.8^{+2.2(4.2)}_{-2.2(4.4)} $  & $-1.92^{+0.07(0.13)}_{-0.07(0.13)} $ & $15.1^{+1.3(2.4)}_{-1.3(2.4)}   $& - & - & 94.9 & 96.5   \\
$y_1^{(4)}$ & $68.3^{+1.8(3.6)}_{-1.8(3.5)} $  & $-0.27^{+0.10(0.19)}_{-0.10(0.20)} $  & $1.92^{+0.64(1.28)}_{-0.64(1.18)} $ & $4.01^{+2.81(6.88)}_{-3.76(6.06)}$  & - & 58.2 & 61.5  \\
$y_1^{(5)}$ & N.D. & N.D. & N.D. & N.D. & N.D. &   N.D. &  N.D. \\
\hline
$y_2^{(3)}$ & $80.3^{+2.4(4.6)}_{-2.4(4.6)} $  & $-2.31^{+0.03(0.05)}_{-0.03(0.05)} $  & $11.23^{+0.37(0.73)}_{-0.37(0.71)}  $ & - & - & 695 & 697  \\
$y_2^{(4)}$ & $77.6^{+1.9(3.6)}_{-1.9(3.6)}$ & $-0.80^{+0.05(0.10)}_{-0.05(0.10)} $ &  $-0.63^{+0.43(0.75)}_{-0.37(0.82)}  $ & $-8.28^{+1.75(2.99)}_{-1.43(3.28)}$ & - & 100 & 103 \\
$y_2^{(5)}$ & $69.9^{+2.2(4.0)}_{-2.2(4.0)}  $ & $-0.54^{+0.13(0.31)}_{-0.18(0.26)} $ & $0.12^{+0.83(1.59)}_{-0.83(1.56)} $ & $4.30^{+0.54(1.35)}_{-0.61(1.17)} $ & $-0.01^{+1.8(3.5)}_{-1.8(3.5)}$ & 2.95 & 7.86 \\
\hline
$P_{2,1}$ & $70.6^{+1.9(3.8)}_{-1.9(3.7)} $ & $-0.56^{+0.11(0.18)}_{-0.09(0.21)} $ & $1.15^{+0.38(1.00)}_{-0.53(0.87)}  $ & - & - & 9.67& 11.3 \\
$P_{2,2}$ & N.C. & N.C. & N.C.& N.C.& - & N.C.& N.C.  \\
$P_{3,2}$ & $70.9^{+2.0(3.9)}_{-2.0(3.8)}$ & $-0.57^{+0.10(0.21)}_{-0.10(0.21)}$ & $1.11^{+0.45(1.10)}_{-0.55(0.96)}$ & $2.12^{+1.51(5.68)}_{-2.62(4.12)}$ & $-0.14^{+1.10(1.83)}_{-1.03(2.09)} $ & 9.39 & 14.3 \\
\hline
\hline
\end{tabular}
\caption{MCMC results at the 68\% (95\%) C.L. for different cosmographic techniques from the combination of low and high-redshift (SNe Ia+OHD+$\mathcal{R}$) data, compared to the derived predictions of the flat $\Lambda$CDM model. N.D. (i.e. \emph{not defined}) indicates that the data are unable to constrain the parameters, whereas N.C. means \emph{no convergence} of the numerical algorithm. $H_0$ values are expressed in km/s/Mpc. The AIC and BIC values are computed with respect to the reference $\Lambda$CDM model.}
\label{tab:low+high z}
\end{center}
\end{table*}

\subsection{Statistical analysis and selection criteria}

Afterwards, we analyze the statistical performances of the various cosmographic approaches through the use of Bayesian selection criteria to measure the evidence of a given model against a reference scenario \citep{Kunz06}, which we chose to be the standard $\Lambda$CDM model. Specifically, we considered the Akaike information criterion (AIC) \citep{Akaike74} and the Bayesian information criterion (BIC) \citep{Schwarz78}, defined as, respectively,
\begin{align}
&\text{AIC}\equiv -2\ln \mathcal{L}_{max}+2p \ , \\
&\text{BIC}\equiv -2\ln \mathcal{L}_{max}+p\ln N\ .
\end{align}
Here, $\mathcal{L}_{max}$ is the maximum likelihood estimate, $p$ is the number of free parameters of the model and $N$ the total number of data points. We note that, for high $N$, the BIC criterion penalizes more severely than AIC the model with a large number of free parameters.

Using these definitions, we calculated the differences $\Delta$AIC and $\Delta$BIC with respect to the corresponding $\Lambda$CDM values to measure the amount of information lost by adding extra parameters in the statistical fitting. Negative values of $\Delta$AIC and $\Delta$BIC indicate that the model under investigation performs better than the reference model, while for positive values, one can refer to:

\begin{itemize}\label{AICintervals}
  \item $\Delta\text{AIC(BIC)}\in [0,2]$ indicates a weak evidence in favour of the reference model, leaving open the question on which model is the most suitable one;
  \item $\Delta\text{AIC(BIC)}\in (2,6]$ indicates a mild evidence against the given model with respect to the reference paradigm;
  \item $\Delta\text{AIC(BIC)}> 6$ indicates a strong evidence against the given model, which should be rejected.
\end{itemize}

\noindent We report the $\Delta$AIC and $\Delta$BIC values for the cosmographic models in the low-redshift regime in \Cref{tab:low z}, and for the combined low and high-redshift regimes in \Cref{tab:low+high z}.

\begin{figure*}
\centering
\subfigure[Third-order analyses\label{subfig24}]{\includegraphics[width=3in]{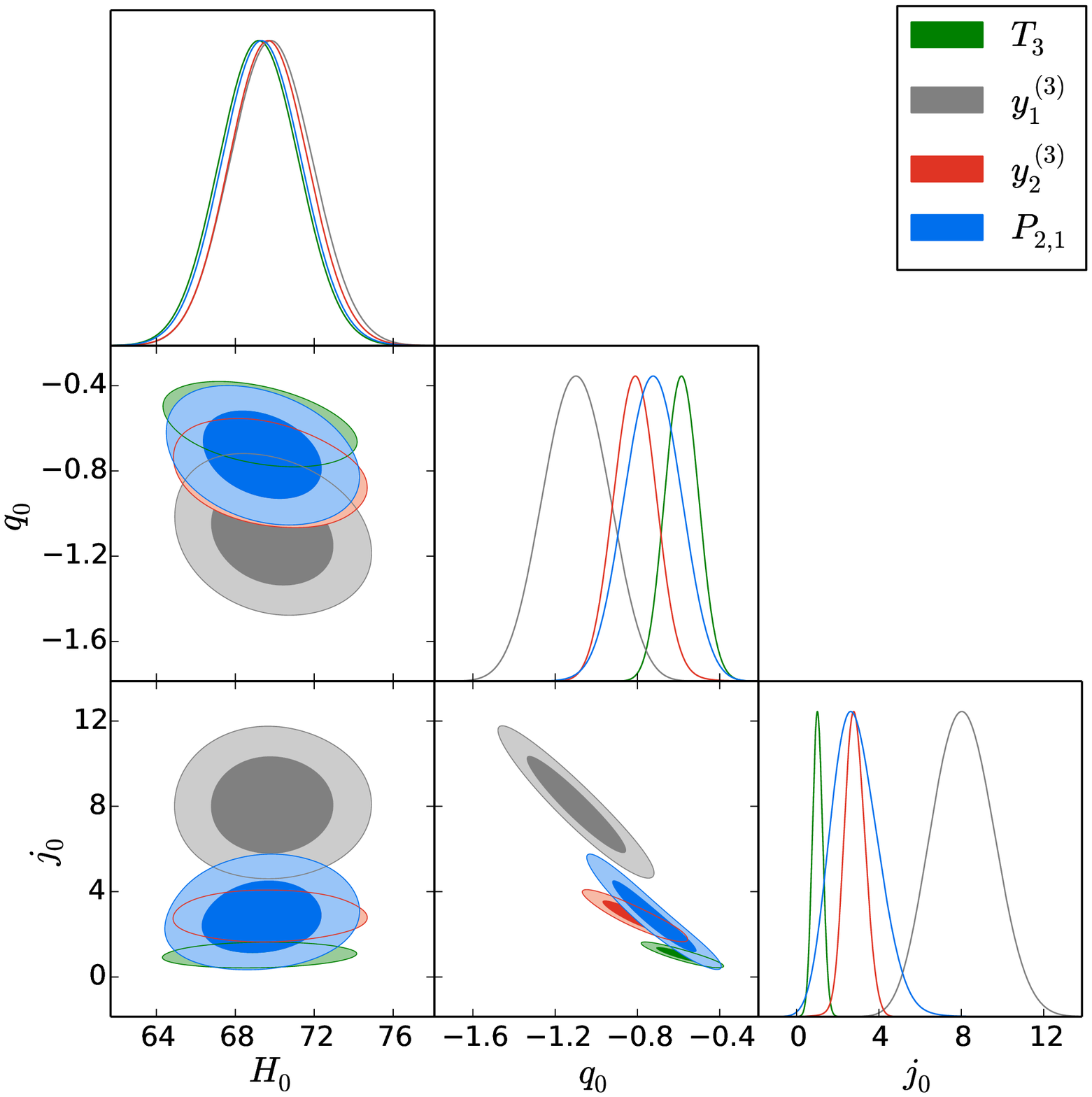}}
\subfigure[Fourth-order analyses\label{subfig24}]{\includegraphics[width=3.8in]{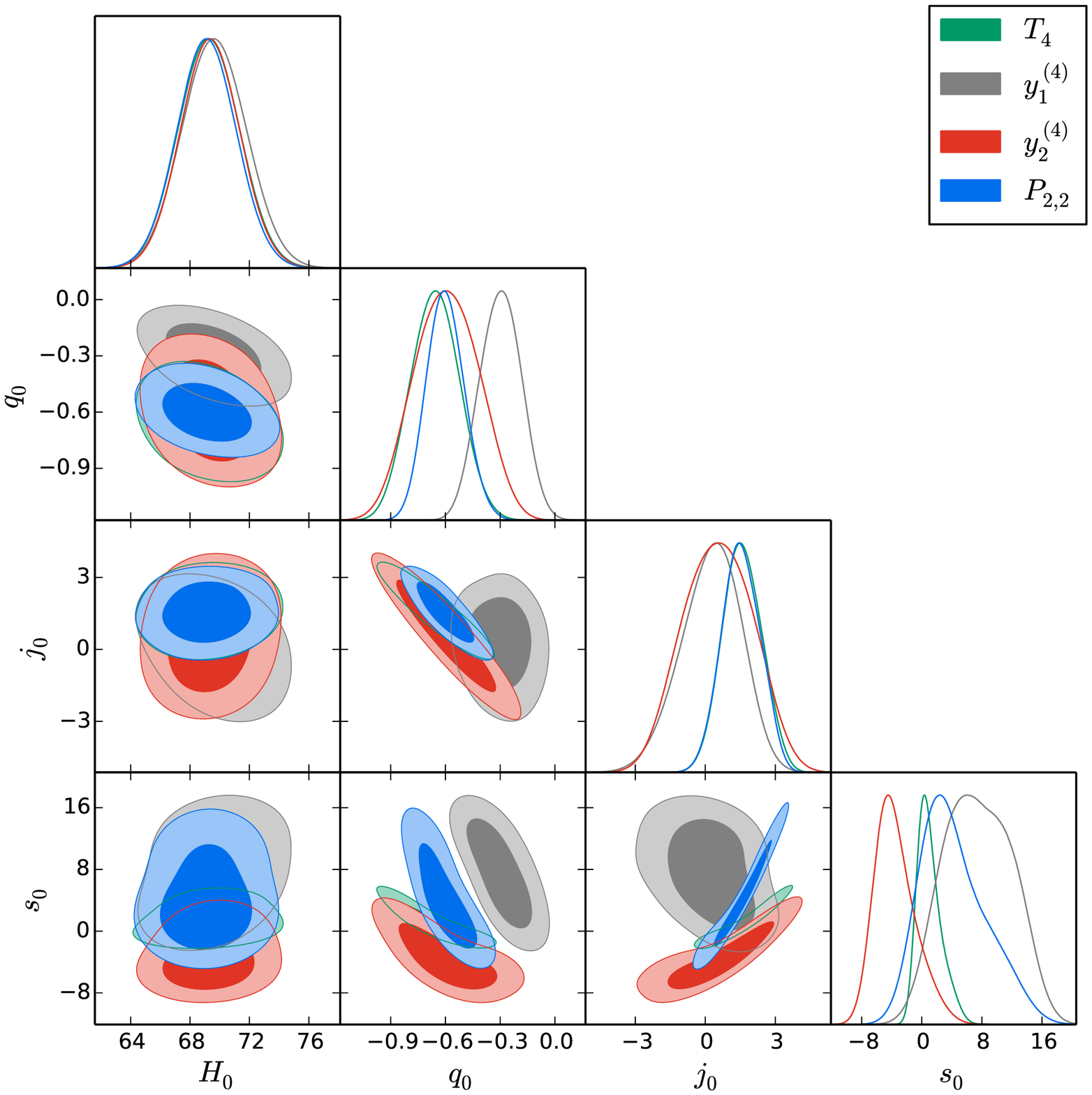}}
\subfigure[Fifth-order analyses\label{subfig24}]{\includegraphics[width=4.2in]{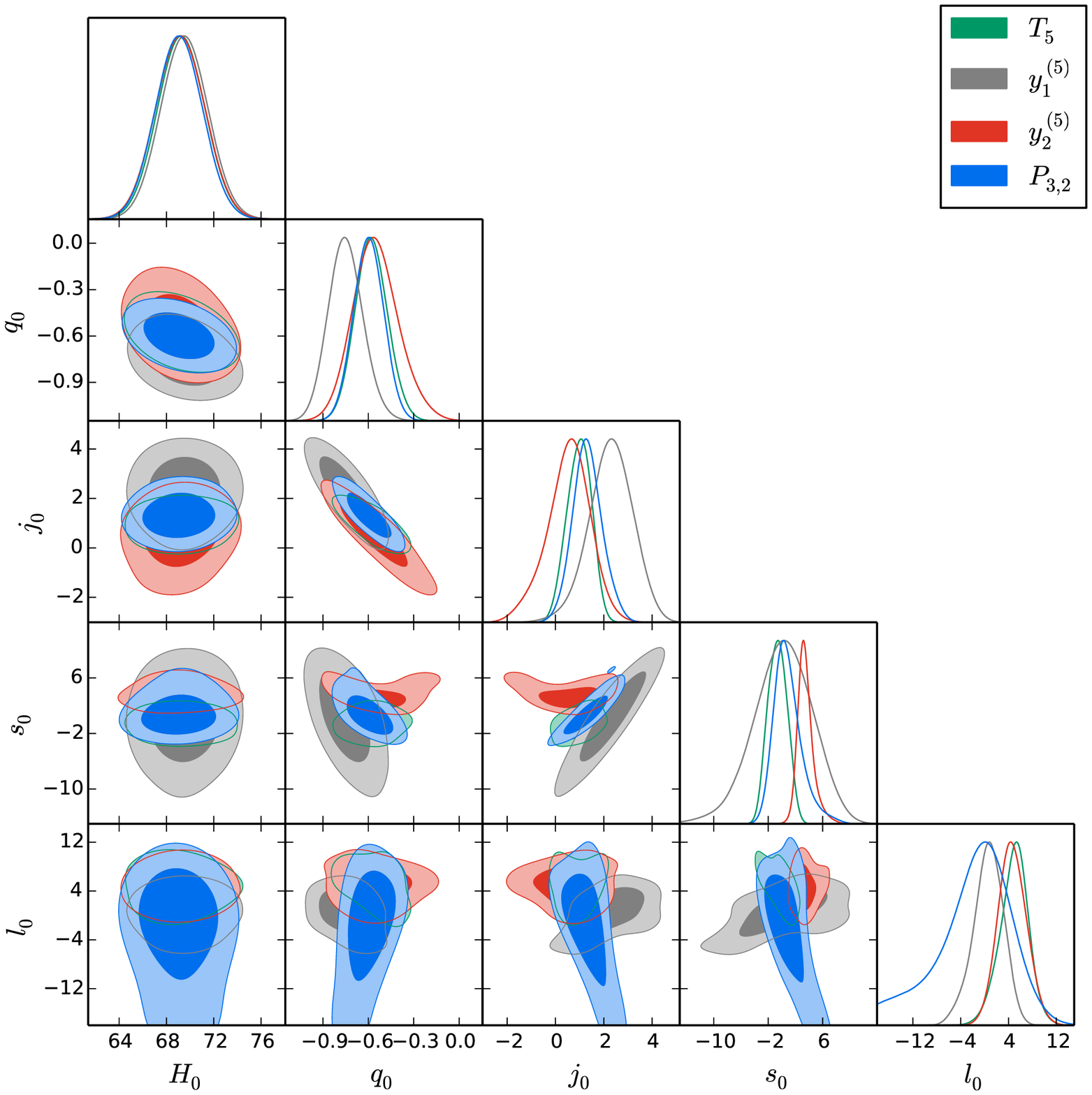}}
\caption{Low-redshift (SNe Ia+OHD) constraints on the cosmographic parameters at the 68\% and 95\% C.L. for different techniques.}
\label{fig:low z}
\end{figure*}

\subsection{Discussion on  numerical results}

Let us now discuss the results we have obtained from the numerical analyses at low and high redshifts. It is worth noticing that the introduction of $y$-variables fails to be predictive and turns out to be a non-viable technique, excepting for a peculiar case built up in terms of $y_2$. In particular, in the low-redshift regime, we note that the results of the third-order $y_1$-variable are substantially different than $2\sigma$ away from the corresponding Taylor constraints. The former results are the tightest and  fully compatible with the $\Lambda$CDM predictions (see top-left panel of \Cref{fig:low z}). However, as suggested by the BIC criterion, the (2,1) Pad\'e polynomial turns out to be the best-performing third-order approximation (cf. \Cref{tab:low z}). At the fourth-order expansion, still the $y_2$-variable behaves better than the $y_1$-variable and provides constraints on the cosmographic parameters that are in agreement with those of the $\Lambda$CDM model within the 68\% C.L. (see top-right panel of \Cref{fig:low z}). The statistical evidence against the  (2,2) Pad\'e polynomial with respect to $\Lambda$CDM is higher compared to the (2,1) Pad\'e polynomial (cf. \Cref{tab:low z}).  At the fifth-order expansion, we clearly note that all the techniques are characterized by large uncertainties on the parameter $l_0$ (see bottom panel of \Cref{fig:low z}). Statistically, the best model is provided by the (3,2) Pad\'e polynomial, whose AIC and BIC values are lower than the corresponding $y$-variables expansions (cf. \Cref{tab:low z}).
Hence, already in the case of small redshifts, the approximations made by
means of Pad\'e polynomials, if accurately calibrated, behave much better in matching the $\Lambda$CDM predictions and in fixing numerical constraints over the cosmographic series. Even the error bars seem to be smaller than the ones obtained from the use of $y$-variables.

This fact is confirmed introducing high-redshift data sets. Indeed, if also the high-redshift measurements are considered, the Taylor polynomials fail to be predictive, as theoretically expected. In this case, the MCMC analysis, at the third-order expansion, indicates that the results of the $y$-variables are many $\sigma$s away from the predictions of the $\Lambda$CDM model, whereas the constraints of the (2,1) Pad\'e polynomial are fully compatible with those of  $\Lambda$CDM  (see top-left panel of \Cref{fig:high z}). These results are confirmed by the AIC and BIC values, which indicate very strong evidence against the $y$-variables (cf. \Cref{tab:low+high z}).
At the fourth-order expansion, the $y$-variables give results in $\sim 3\sigma$ tension with each other (see top-right panel of \Cref{fig:high z}), and values of the set $(q_0,s_0,l_0)$ are, in both cases, $\gtrsim 2\sigma$ away from those predicted by the $\Lambda$CDM model (cf. \Cref{tab:low+high z}). On the other hand, the (2,2) Pad\'e rational polynomial does not represent a suitable approximation of $d_L(z)$ in the high-redshift domains, where it shows a plateau behaviour (see \Cref{fig:Pade_curves}) due to the presence of the same powers of $z$ in the numerator and denominator. At the fifth-order expansion, the cosmographic series, provided by the $y_1$-variable, is not constrained by the data, while the  $y_2$-variable is characterized by the lowest AIC and BIC values among the different techniques of the same order (cf. \Cref{tab:low+high z}). The results of the (3,2) Pad\'e polynomial are in agreement with the predictions of the $\Lambda$CDM model at the 68\% C.L., although the large uncertainties on the high-order cosmographic parameters $s_0$ and $l_0$ (see bottom panel of \Cref{fig:high z}). See also \cite{lusso} for a discussion on $\Lambda$CDM tension.

The previous considerations suggest that model-independent \emph{geometrical} tests could be relevant to check the goodness of our findings. In particular, they are essential to understand if the concordance paradigm is effectively recovered in the picture of our numerical results. To this end, the use of statefinder diagnostic is of utmost importance because these quantities depend on space-time geometry only\footnote{This is a consequence of the fact that they are functions of Hubble rate or scale factor.} \citep{refer1}. The statefinder prescription will be used in the next subsection, motivated by the fact that, at both $1$ and $2\sigma$ confidence levels, using low and high redshift data, the $\Lambda$CDM predictions are not completely fulfilled by our analyses. For example, the most \emph{stable} techniques, i.e. the auxiliary variables and Pad\'e polynomials that minimize the AIC and BIC values, suggest that the jerk parameter seems to be slightly larger than the one predicted by the concordance paradigm, i.e. $j_0=1$. We discuss in detail this result in the next subsection.

To summarize, the $y_2$-parametrization performs relatively better than the $y_1$-parametrization over the whole redshift domain, while the most accurate and stable approximations of $d_L(z)$, at the high-redshift regimes,  are given by the Pad\'e technique. Given the low number of degrees of freedom and its ability to properly constrain the cosmographic series, the (2,1) Pad\'e approximation is the best polynomial to be used in high-redshift cosmography. The full list of results is summarized in \Cref{tab:sommario}.

\begin{table*}
\small
\begin{tabularx}{\textwidth}{p{4cm} p{5cm} p{5cm}}
\centering Cosmographic technique & \centering Low redshifts & \hspace{1.5cm}High redshifts  \\
\hline
\begin{center} Taylor  \end{center} &
\begin{itemize}
\item[$\blacktriangleright$] The series is suitable when data are inside $z<1$.
\item[$\blacktriangleright$] Good agreement with the $\Lambda$CDM model, up to the fourth order.
\item[$\blacktriangleright$] Limited to short intervals of data.
\end{itemize} &
\begin{itemize}
\item[$\blacktriangleright$] High-redshift measurements are non-predictive.
\item[$\blacktriangleright$] Numerical outcomes are both non-physical and non-convergent.
\item[$\blacktriangleright$] Useless for making physical predictions.
\end{itemize}\\
\hline
\begin{center}$y$-variable  \end{center} &
\begin{itemize}
\item[$\blacktriangleright$] Third-order $y_1$ is unsuitable as more than $2\sigma$ away from Taylor cosmography.
\item[$\blacktriangleright$] At the fourth order, $y_2$ works much better than $y_1$.
\item[$\blacktriangleright$] No apparent need to adopt $y$-variables instead of Taylor series.
\end{itemize} &
\begin{itemize}
\item[$\blacktriangleright$] At all orders, $y_1$ is unsuitable.
\item[$\blacktriangleright$] Only the fifth-order $y_2$ is suitable and shows the lowest AIC and BIC values among all the fits.
\item[$\blacktriangleright$] $\Delta $AIC and $\Delta $BIC  indicate very strong evidence against the $y$-variables, except for the fifth-order $y_2$.
\end{itemize}\\
\hline
\begin{center} Pad\'e \end{center}&
\begin{itemize}
\item[$\blacktriangleright$] The (2,1) polynomial is statistically the best-performing approximation.
\item[$\blacktriangleright$]  The (2,2) and (3,2) polynomials give similar results, but with unsuitable errors.
\item[$\blacktriangleright$] No apparent need to use Pad\'e series of any orders over the Taylor series.
\end{itemize} &
\begin{itemize}
\item[$\blacktriangleright$] The (2,2) Pad\'e polynomial is strongly disfavoured and should be ruled out.
\item[$\blacktriangleright$] The (2,1) and (3,2) polynomials are suitable approximations, but with very large uncertainties.
\item[$\blacktriangleright$] The (2,1) polynomial performs better than the $y$-variables.
\end{itemize}\\
\hline
\end{tabularx}
\caption{Comparison among the different cosmographic techniques.}
\label{tab:sommario}
\end{table*}


\subsection{Discussion on a possible dark energy evolution}

In the previous subsection, we left open the possibility that dark energy could slightly evolve with time. This is plausible at both $1\sigma$ and $2\sigma$ confidence levels on the basis of our numerical outcomes. Let us now investigate better this fact through the use of the well established statefinder diagnostics. In particular the statefinder parameter is defined by
\begin{equation}\label{statefy}
Omh^2(z_i;z_j) = \frac{h^2(z_i) - h^2(z_j)}{(1+z_i)^3 - (1+z_j)^3}\,,
\end{equation}
where $h(z) = H(z)/100$km/sec/Mpc and $z_i,z_j$ represent a set of two redshifts at which the Hubble rates of Eq. \eqref{statefy} are computed. Concerning the concordance paradigm, one immediately has $ Omh^2 = \Omega_{m0}\,h^2$ and looking at the most recent Planck constraint, we can immediately deduce that
\begin{equation}\label{plaomh2}
Omh^2=\Omega_{m0}\,h^2=0.14240 \pm 0.00087\,.
\end{equation}
This value is in tension with the $\Lambda$CDM-based prediction, stable around $\Omega_{m0}\,h^2\simeq0.12$, when Baryonic Acoustic Oscillation measurements are considered (see for details \citet{refer3,refer2}).

To go further and checking this with cosmography, it is possible to reproduce Eq. \eqref{plaomh2}, starting with the recipe that a generic dark energy model evolves as $H(z)=H_0\sqrt{\Omega_{m0}(1+z)^{3}+\Omega_{DE0}G(z)}$, where $G(z)$ satisfies $G(z)\rightarrow 1$, as $z=0$, with  $\Omega_{DE0}\equiv1-\Omega_{m0}$. Thus, we can write
\begin{equation}\label{qudef}
q(z)=-1+\frac{(1+z)\left[3\Omega_{m0}(1+z)^{2}+\Omega_{DE0}G'(z)\right]}{2\left[\Omega_{m0}(1+z)^{3}+\Omega_{DE0}G(z)\right]}\,,
\end{equation}
that, for a genuine cosmological constant contribution, becomes in terms of cosmography:
\begin{equation}\label{q0nonevolving}
Omh^2=\frac{2}{3}(q_0+1)h^2\,.
\end{equation}
Taking our numerics, it is immediately possible to find the statefinder values at small and high redshift domains by involving $h=h_0$ and $q_0$ as reported in Tabs. I and II. Thus, at small redshifts, i.e. using only SNe Ia and OHD data sets, we have
\begin{eqnarray}\label{calcul1}
Omh^2_{y, small}   &=& 0.1316\pm0.0618(0.1188)\,, \\
Omh^2_{P, small}   &=& 0.0864\pm0.0466(0.0928)\,,
\end{eqnarray}
in which we evaluated, through Eq. \eqref{q0nonevolving}, the \emph{best} auxiliary variable approximation, i.e. $y_2^{(4)}$, and the \emph{optimal} rational approximation, i.e. $P_{21}$, respectively.

\noindent The same procedure can be considered at higher redshifts, where we have instead
\begin{eqnarray}\label{calcul2}
Om\,h^2_{y, high}   &=& 0.1498\pm0.0599(0.1100)\,, \\
Om\,h^2_{P, high}   &=& 0.1441\pm0.0416(0.0860)\,,
\end{eqnarray}
for the most suitable auxiliary variable, this time $y_2^{(5)}$, and the optimal version of Pad\'e approximation, given now by $P_{3,2}$,  respectively. We explored only these four scenarios, because these approximations are the statistically favorite ones, as confirmed by AIC and BIC statistical criteria, as one can see in Tabs. I and II.

At small redshifts only, the tensions among our predictions and the Planck measurements still persist, becoming larger in the rational approximation case, i.e. for Pad\'e (2,1). This can be interpreted as previously stated throughout the manuscript: \emph{the use of rational approximations works better only when the redshift is large} and becomes unnecessary at large scales, i.e. when the redshift becomes smaller. At larger redshifts, however, adding the $\mathcal R$ measure, the tension between our $\Lambda$CDM predictions and the Planck results, is alleviated. As a consequence, the $\Lambda$CDM model can be \emph{recovered} somehow, albeit in both cases the error bars are quite large at $1\sigma$ confidence level yet, leaving open the possibility that $Omh^2$ persists being in tension\footnote{For the sake of completeness, we notice that we computed our error bars through a standard logarithmic formula. As errors over $q_0$ and $h_0$, we considered the mean values among upper and lower error limits.}. So, even in this case a dark energy evolution is not excluded \emph{a priori}. To justify our prescriptions, one can imagine that possible systematics can affect the whole analysis in general, without substantially alleviating the tension presented in \cite{refer2}. This limitation has been discussed for example in \cite{aert}.

In addition, for generic dark energy models, inverting Eq. \eqref{qudef}, cosmography suggests that
\begin{equation}\label{q0evolving}
\Omega_{m0}=\frac{G_0^\prime-2(q_0+1)}{G_0^\prime-3}\,.
\end{equation}
Moreover, by virtue of the dark energy form, one immediately finds
\begin{equation}\label{jeidef}
\frac{G_0^{''}}{3-G_0^\prime}\geq\frac{2G_0^\prime}{3-G_0^\prime}\,,
\end{equation}
which has been evaluated assuming $j_0\geq1$, as our numerical results partially seem to indicate, and conventionally considering that $G^\prime(0)\equiv G_0^\prime$ and $G''(0)\equiv G_0^{''}$. From the above considerations, cosmographic departures of $Omh^2$, above depicted, can represent a signal that $\Lambda$ does not drive the universe to accelerate today\footnote{We note that the denominator sign of Eq. \eqref{jeidef} is not specified \emph{a priori}. Thence, if positive, one has $G_0^{''}>2G_0^\prime$, otherwise $G_0^{''}<-2\big|G_0^\prime\big|$.}. This would imply that some sort of modifications of Einstein's gravity is still possible. By looking at Eqs. \eqref{q0evolving} and \eqref{jeidef}, models in which dark energy, and in particular the cosmological constant, is {\em screened} (or compensated) by a dynamically evolving counter-term are permitted, once $G_0^\prime\neq3$, choosing the sign of both $G_0^\prime$ and $G_0''$. Concluding, the dark energy evolution is not excluded \emph{a priori} through the here-developed cosmographic methods. A possible tension among $\Omega_{m0}\,h^2$ measured by Planck and expected by cosmography can lead to the need of new physics. A more accurate study on these aspects will be the object of future efforts toward the refinements of a new \emph{high-redshift cosmography}. This would characterize the signs of $G_0^\prime$ and $G_0^{''}$ with increasing accuracy, disclosing at small and large redshifts whether dark energy evolves with time.


\section{Final outlooks and perspectives}
\label{sec:sezione5}

We  critically revised the role of cosmographic treatments and, in particular, the reconstruction techniques widely adopted in the literature. The main issues of cosmography have been considered  focusing on the basic demands of the convergence problem, related to series expansions around $z=0$, while cosmic data go further this limit. We first reviewed the two main scenarios introduced to heal the convergence issue, i.e. the rational approximations and the auxiliary variables, showing how to construct the Pad\'e  and the $y$-variable cosmography, respectively. From a theoretical point of view, we developed the main rules and the basic requirements which any reconstruction should have. An important result is  that  lower order might be favored than higher ones. The theoretical issues associated with the use of $y$-variables have been also discussed. We took into account a hierarchy between cosmographic coefficients, starting from the third-order expansion up to the fifth one.

Although appealing, $y$-variable cosmography becomes non predictive already from a theoretical point of view. Thus, in agreement with previous approaches, we showed that parametrizations of the redshift variables and alternatives to Taylor expansions are disfavoured if used arbitrarily, i.e. without calibrating the orders with cosmic data. Specifically, we considered the two parametrizations $y_1=1-a$ and $y_2=\arctan(a^{-1}-1)$, and some Pad\'e expansions. It is worth noticing  that approximations different from $y_1$ and $y_2$,  diverging for   $z\rightarrow\infty$, are not theoretically viable to heal the convergence issue. Afterwards, we focused on the (1,1), (2,1), (2,2), (3,2) and (4,1) Pad\'e approximations, discarding  other orders which  show bad behaviours if compared to the fiducial $\Lambda$CDM model. We therefore provided numerical fits by means of Monte Carlo integration of combined Supernova Ia data and Hubble measurements. Specifically, we adopted the most recent Pantheon data set, free from nuisance parameters due to the adopted condition of a spatially flat universe, and the observational Hubble data acquired through the differential age technique. We tested different cosmographic orders by splitting our study at late times, and early times through the use of the CMB shift parameter. As theoretically argued, we found non conclusive or non convergent results in several cases of interest, erroneously used in the literature. Moreover,  the AIC and BIC selection criteria have been adopted as tools for inferring the statistical significance of a given scenario with respect to the reference $\Lambda$CDM model.

In particular, in the low-redshift regime, we noted that the results of the third-order $y_1$-variable are substantially different than $2\sigma$ with respect to the corresponding Taylor constraints, whereas the (2,1) Pad\'e polynomial turns out to be the best-performing third-order approximation. Going further with the orders, still the $y_2$-variable behaves better than the $y_1$-variable, while a strong statistical evidence against the  (2,2) Pad\'e polynomial is found. At the fifth-order expansion, large uncertainties occur, making cosmography non predictive, as expected.
The value of $l_0$ results to be unbounded in every analysis, and a statistical significance seems to favor the (2,1) Pad\'e polynomial over the (3,2) one, although the interesting results provided by the latter.

Involving high-redshift measurements, we found that every order of Taylor polynomials fails to be predictive, as theoretically expected. Even the $y$-variables are many $\sigma$s away from the predictions of the $\Lambda$CDM model, whereas the constraints of the (2,1) Pad\'e polynomial are fully compatible with those ones.
Increasing the order leads to more problematic results, as the fourth-order $y$-variables give results that are in $\sim 3\sigma$ tension with each other, and the set $(q_0,s_0,l_0)$ is, in both cases, in $\gtrsim 2\sigma$  tension with the predictions of the $\Lambda$CDM model.
The (2,2) Pad\'e rational polynomial does not represent a suitable approximation of $d_L(z)$ at high redshifts, due to the presence of the same powers of $z$ in the numerator and denominator. The main troubles occur as soon as the fifth-order expansions are involved. The cosmographic series provided by the $y_1$-variable is not constrained by the data, while we found that the $y_2$ variable works better than any other expansions.
Our results show that the $y_2$-parametrization performs relatively better than the $y_1$-parametrization over the whole redshift domains, while the most accurate and stable approximations of $d_L(z)$ at the high-redshift regimes are due to the Pad\'e approximations. The Bayesian information criteria suggest that the (2,1) Pad\'e approximation is optimal, also due to the low number of variables involved in the approximations. A tension with the concordance paradigm was also found by investigating the statefinder diagnostics. In particular, comparing our results with the ones got by Planck, we showed that the $\Lambda$CDM model seems to predict smaller values of $Omh^2$, in net tension with Planck results. This is particularly true at small redshifts, whereas by adding the $\mathcal R$ measure, at smaller scales and higher redshifts, it is weakly alleviated, albeit error bars do not permit to conclude that the tension is removed. Indeed, it seems that both cases leave open the possibility that dark energy evolves in time.

Future works will focus on how to generalize the aforementioned results by also considering alternative rational approximations.  We will focus on how to reduce systematics, minimizing the expansion orders and maximizing the quality of numerical outcomes up to the $l_0$ order. The same procedure, reported here, will be considered in view of the whole Planck data to check whether cosmography is able to discriminate the evolution of dark energy from the cosmological constant evolution.

As a final remark, it is worth saying that the
cosmographic approach  can  provide  a useful feedback  on the behaviour of dark energy, in particular  at large redshifts of the order $z\sim 10$.  This  is valuable in order to see how the
late-time matter domination exploding (oscillating) terms (for example in modified gravity) may (or may not) affect the equation of state of dark energy. This will  be the topic of a forthcoming paper.

\section*{Acknowledgements}
The authors acknowledge the support of INFN (iniziative specifiche QGSKY and MoonLIGHT-2). This paper is partially based upon work from COST action CA15117 (CANTATA), supported by COST (European Cooperation in Science and Technology). O.L. acknowledges the support provided by the Ministry of Education and Science of the Republic of Kazakhstan, Program IRN: BR05236494.

\appendix

\onecolumn

\section*{Appendix}
\label{appendix}

Here, all the cosmographic expansions, up to the fifth order, which we adopted in this study are reported.
\\
\begin{itemize}
\item Taylor approximations of the luminosity distance and the Hubble rate:

\begin{align}
d_L(z)=&\ \dfrac{1}{H_0}\Big[z +  \dfrac{1}{2}(1 - q_0) z^2 - \dfrac{1}{6}(1 - q_0 - 3 q_0^2 + j_0) z^3+ \dfrac{1}{24}(2 - 2 q_0 - 15 q_0^2 - 15 q_0^3 + 5 j_0 \nonumber  \\
&+10 q_0 j_0 +s_0) z^4 +  \Big(-\dfrac{1}{20} - \dfrac{9j_0 }{40}+ \dfrac{j_0^2}{12} - \dfrac{l_0}{120} + \dfrac{q_0}{20} - \dfrac{11 j_0 q_0}{12} + \dfrac{27 q_0^2}{40} - \dfrac{7 j_0 q_0^2}{8}  \nonumber  \\
&+ \dfrac{11 q_0^3}{8} + \dfrac{7 q_0^4}{8} - \dfrac{11 s_0}{120} -\dfrac{q_0 s_0}{8}\Big) z^5\Big]\ ,
\end{align}
\vspace{-0.2cm}
\begin{align}
H(z)=&\ H_0\Big[1 + (1 + q_0) z + \dfrac{1}{2} (j_0 - q_0^2) z^2 - \dfrac{1}{6}(-3 q_0^2 - 3 q_0^3 + j_0 (3 + 4 q_0) + s_0) z^3  \nonumber \\
 & +\dfrac{1}{24}(-4 j_0^2 + l_0 - 12 q_0^2 - 24 q_0^3 - 15 q_0^4 + j_0 (12 + 32 q_0 + 25 q_0^2) + 8 s_0 + 7 q_0 s_0) z^4\Big]\ .
\end{align}
\\
\item  $y_1$-variable approximations of the luminosity distance and the Hubble rate:

\begin{align}
d_L(y_1)=&\ \dfrac{1}{H_0}\Big[y_1 + \dfrac{1}{2} (3- q_0) y_1^2 + \dfrac{1}{6} (11 - j_0 - 5 q_0 + 3 q_0^2) y_1^3 + \dfrac{1}{24} (50 - 7 j_0 - 26 q_0 + 10 j_0 q_0\nonumber \\
& + 21 q_0^2 - 15 q_0^3 + s_0) y_1^4 + \dfrac{1}{120} (274 - 47 j_0 + 10 j_0^2 - l_0 - 154 q_0 + 90 j_0 q_0 + 141 q_0^2 \nonumber \\
&- 105 j_0 q_0^2 - 135 q_0^3 + 105 q_0^4 + 9 s_0 - 15 q_0 s_0) y_1^5\Big]\ ,
\end{align}
\vspace{-0.2cm}
\begin{align}
H(y_1)=&\ H_0\Big[1 + (1 + q_0) y_1 + \dfrac{1}{2} (2 + j_0 + 2 q_0 - q_0^2) y_1^2 - \dfrac{1}{6}(-6 - 6 q_0 + 3 q_0^2 - 3 q_0^3 \nonumber \\
&+ j_0 (-3 + 4 q_0) + s_0) y_1^3-\dfrac{1}{24} (-24 + 4 j_0^2 - l_0 - 24 q_0 + 12 q_0^2 - 12 q_0^3 + 15 q_0^4 \nonumber \\
&+  j_0 (-12 + 16 q_0 - 25 q_0^2) + 4 s_0 - 7 q_0 s_0) y_1^4\Big]\ .
\end{align}
\\
\item $y_2$-variable approximations of the luminosity distance and the Hubble rate:

\begin{align}
d_L(y_2)=&\ \dfrac{1}{H_0}\Big[y_2 + \dfrac{1}{2}(1 - q_0) y_2^2 + \dfrac{1}{6} (1 - j_0 + q_0 + 3 q_0^2) y_2^3+\dfrac{1}{24}(10 + 5 j_0 - 10 q_0 + 10 j_0 q_0  \nonumber \\
& - 15 q_0^2- 15 q_0^3 + s_0) y_2^4+\dfrac{1}{120} (-10 - 47 j_0 + 10 j_0^2 - l_0 + 26 q_0 - 110 j_0 q_0 + 141 q_0^2 \nonumber \\
 &- 105 j_0 q_0^2 + 165 q_0^3 + 105 q_0^4 - 11 s_0 - 15 q_0 s_0) y_2^5\Big]\ ,
\end{align}
\vspace{-0.2cm}
\begin{align}
H(y_2)=&\ H_0\Big[1 + (1 + q_0) y_2 +\dfrac{1}{2} (j_0 - q_0^2) y_2^2 + \dfrac{1}{6} (2 + 2 q_0 + 3 q_0^2 + 3 q_0^3 - j_0 (3 + 4 q_0) - s_0) y_2^3 \nonumber \\
& +\dfrac{1}{24}(-4 j_0^2 + l_0 - 20 q_0^2 - 24 q_0^3 - 15 q_0^4 +  j_0 (20 + 32 q_0 + 25 q_0^2) + 8 s_0 + 7 q_0 s_0) y_2^4 \Big]\ .
\end{align}
\\
\item (2,1) Pad\'e approximation of the luminosity distance:

\begin{equation}
P_{2,1}(z)=\dfrac{1}{H_0}\left[\dfrac{z (6 (-1 + q_0) + (-5 - 2 j_0 + q_0 (8 + 3 q_0)) z)}{-2 (3 + z + j_0 z) + 2 q_0 (3 + z + 3 q_0 z)}\right] .
\end{equation}
\\
\item (2,2) Pad\'e approximation of the luminosity distance:

\begin{align}
P_{2,2}(z)&=\dfrac{1}{H_0}(6 z (10 + 9 z - 6 q_0^3 z + s_0 z - 2 q_0^2 (3 + 7 z) - q_0 (16 + 19 z) +
j_0 (4 + (9 + 6 q_0) z))\Big/ \nonumber \\	
&(60 + 24 z + 6 s_0 z - 2 z^2	+ 4 j_0^2 z^2 - 9 q_0^4 z^2 - 3 s_0 z^2 + 6 q_0^3 z (-9 + 4 z) + q_0^2 (-36 - 114 z \nonumber \\
&+ 19 z^2)  +j_0 (24 + 6 (7 + 8 q_0) z + (-7 - 23 q_0 + 6 q_0^2) z^2) +  q_0 (-96 - 36 z + (4 + 3 s_0) z^2))\ .
\end{align}

\item (3,2) Pad\'e approximation of the luminosity distance:

\begin{align}
P_{3,2}(z)&=\dfrac{1}{H_0}(z (-120 - 180 s_0 - 156 z - 36 l_0 z - 426 s_0 z - 40 z^2 + 80 j_0^3 z^2 - 30 l_0 z^2 - 135 q_0^6 z^2  \nonumber \\
    &-210 s_0 z^2 + 15 s_0^2 z^2 - 270 q_0^5 z (3 + 4 z) + 9 q_0^4 (-60 + 50 z + 63 z^2) + 2 q_0^3 (720 + 1767 z \nonumber \\
    &+ 887 z^2) + 3 j_0^2 (80 + 20 (13 + 2 q_0) z + (177 + 40 q_0 - 60 q_0^2) z^2) + 6 q_0^2 (190 + 5 (67 + 9 s_0) z  \nonumber \\
    &+(125 + 3 l_0 + 58 s_0) z^2) -6 q_0 (s_0 (-30 + 4 z + 17 z^2) - 2 (20 + (31 + 3 l_0) z + (9 + 4 l_0) z^2)) \nonumber \\
    &+6 j_0 (-70 + (-127 + 10 s_0) z + 45 q_0^4 z^2 + (-47 - 2 l_0 + 13 s_0) z^2 + 5 q_0^3 z (30 + 41 z) \nonumber \\
    & - 3 q_0^2 (-20 + 75 z + 69 z^2) + 2 q_0 (-115 - 274 z + (-136 + 5 s_0) z^2))))\Big/(3(-40 - 60 s_0 \nonumber \\
    &-32 z - 12 l_0 z - 112 s_0 z - 4 z^2 + 40 j_0^3 z^2 - 4 l_0 z^2 - 135 q_0^6 z^2 - 24 s_0 z^2 + 5 s_0^2 z^2 \nonumber \\
    &- 30 q_0^5 z (12 + 5 z) + 3 q_0^4 (-60 + 160 z + 71 z^2) +j_0^2 (80 + 20 (11 + 4 q_0) z + (57 + 20 q_0 \nonumber \\
    &- 40 q_0^2) z^2) + 6 q_0^3 (80 + 188 z + (44 + 5 s_0) z^2) + 2 q_0^2 (190 + 20 (13 + 3 s_0) z + (46 + 6 l_0 \nonumber \\
    &+ 21 s_0) z^2)+4 q_0 (20 + (16 + 3 l_0) z + (2 + l_0) z^2 + s_0 (15 - 17 z - 9 z^2))+2 j_0 (-70 \nonumber \\
    &+ 2 (-46 + 5 s_0) z + 90 q_0^4 z^2 + (-16 - 2 l_0 + 3 s_0) z^2 + 15 q_0^3 z (12 + 5 z) + q_0^2 (60 - 370 z \nonumber \\
    &- 141 z^2) + 2 q_0 (-115 - 234 z + 2 (-26 + 5 s_0) z^2))))\ .
\end{align}

\end{itemize}

\bsp	
\label{lastpage}

\end{document}